\documentclass[twocolumn]{aastex63}

\usepackage{amsmath}
\usepackage{gensymb}
\usepackage{booktabs}
\usepackage{xcolor}
\usepackage[ruled, vlined, linesnumbered]{algorithm2e}
\usepackage{algpseudocode}
\usepackage{fontawesome5}
\usepackage{hyperref}

\def\ltsima{$\; \buildrel < \over \sim \;$}
\def\simlt{\lower.5ex\hbox{\ltsima}}
\def\gtsima{$\; \buildrel > \over \sim \;$}
\def\simgt{\lower.5ex\hbox{\gtsima}}
\def\pc{{\rm\,pc}}

\def\g{{\rm\,g}}

\def \r {r$^{1/4}$ }


\def\s{\ifmmode \widetilde \else \~\fi}
\def\={\overline}

\def\spose#1{\hbox to 0pt{#1\hss}}

\def\eg{{e.g.,\ }}
\def\ie{{i.e.\ }}
\def\lta{\mathrel{\spose{\lower 3pt\hbox{$\mathchar"218$}}
     \raise 2.0pt\hbox{$\mathchar"13C$}}}
\def\gta{\mathrel{\spose{\lower 3pt\hbox{$\mathchar"218$}}
     \raise 2.0pt\hbox{$\mathchar"13E$}}}
\def\Dt{\spose{\raise 1.5ex\hbox{\hskip3pt$\mathchar"201$}}}    
\def\dt{\spose{\raise 1.0ex\hbox{\hskip2pt$\mathchar"201$}}}    

\def\r{${\rm r^{1/4}}$}

\def\dotsfill{\leaders\hbox to 1em{\hss.\hss}\hfill}
\def\sun{\odot}
\def\earth{\oplus}

\def\ms{{\rm\,ms}}

\def\ltsima{$\; \buildrel < \over \sim \;$}
\def\gtsima{$\; \buildrel > \over \sim \;$}
\def\lsim{\lower.5ex\hbox{\ltsima}}
\def\gsim{\lower.5ex\hbox{\gtsima}}
\def\lapp{\ifmmode\stackrel{<}{_{\sim}}\else$\stackrel{<}{_{\sim}}$\fi}
\def\gapp{\ifmmode\stackrel{>}{_{\sim}}\else$\stackrel{<}{_{\sim}}$\fi}

\definecolor{myred}{HTML}{c00028}
\definecolor{mydarkblue}{HTML}{005353}
\definecolor{myddblue}{HTML}{048080}
\definecolor{myblue}{HTML}{098f94}

\defcitealias{V21}{V21}
\usepackage{subfigure}

\newcommand{\github}[1]{%
   \href{#1}{\textcolor{black}\faGithubSquare}%
}

\submitjournal{ApJ}
\received{March 03, 2023}
\revised{October 06 2023}
\accepted{October 06, 2023}

\shorttitle{Physical Symbolic Optimization}

\shortauthors{Tenachi et al.}
\graphicspath{{./}{figures/}}

\begin{document}

\title{Deep symbolic regression for physics guided by units constraints:\\ toward the automated discovery of physical laws}

\def\PhySO{{$\Phi$-SO}}
\def\insitu{{\textit{in situ}\ }}
\def\Insitu{{\textit{In situ}\ }}
\def\posthoc{{\textit{post hoc}\ }}
\def\Posthoc{{\textit{Post hoc}\ }}
\def\placeholder{{\square}}


\author[0000-0001-8392-3836]{Wassim Tenachi}
\affiliation{Universit\'e de Strasbourg, CNRS, Observatoire astronomique de Strasbourg, UMR 7550, F-67000 Strasbourg, France}

\author[0000-0002-3292-9709]{Rodrigo Ibata}
\affiliation{Universit\'e de Strasbourg, CNRS, Observatoire astronomique de Strasbourg, UMR 7550, F-67000 Strasbourg, France}

\author[0000-0002-8788-8174]{Foivos I. Diakogiannis}
\affiliation{Data61, CSIRO, Kensington, WA 6155, Australia}

\begin{abstract}
Symbolic Regression is the study of algorithms that automate the search for analytic expressions that fit data. While recent advances in deep learning have generated renewed interest in such approaches, the development of symbolic regression methods has not been focused on physics, where we have important additional constraints due to the units associated with our data. Here we present $\Phi$-SO, a Physical Symbolic Optimization framework for recovering analytical symbolic expressions from physics data using deep reinforcement learning techniques by learning units constraints. Our system is built, from the ground up, to propose solutions where the physical units are consistent by construction. This is useful not only in eliminating physically impossible solutions, but because the ``grammatical'' rules of dimensional analysis restrict enormously the freedom of the equation generator, thus vastly improving performance. The algorithm can be used to fit noiseless data, which can be useful for instance when attempting to derive an analytical property of a physical model, and it can also be used to obtain analytical approximations to noisy data. We test our machinery on a standard benchmark of equations from the Feynman Lectures on Physics and other physics textbooks, achieving state-of-the-art performance in the presence of noise (exceeding 0.1\%) and show that it is robust even in the presence of substantial (10 \%) noise. We showcase its abilities on a panel of examples from astrophysics.
\end{abstract}
\keywords{Symbolic regression, Reinforcement learning, Recurrent neural network, Dimensional analysis, Physical units, Grammar guided symbolic generation}

\section{Introduction}
\label{sec:Introduction}

\begin{figure*}
\begin{center}
\includegraphics[angle=0, clip, width=\hsize]{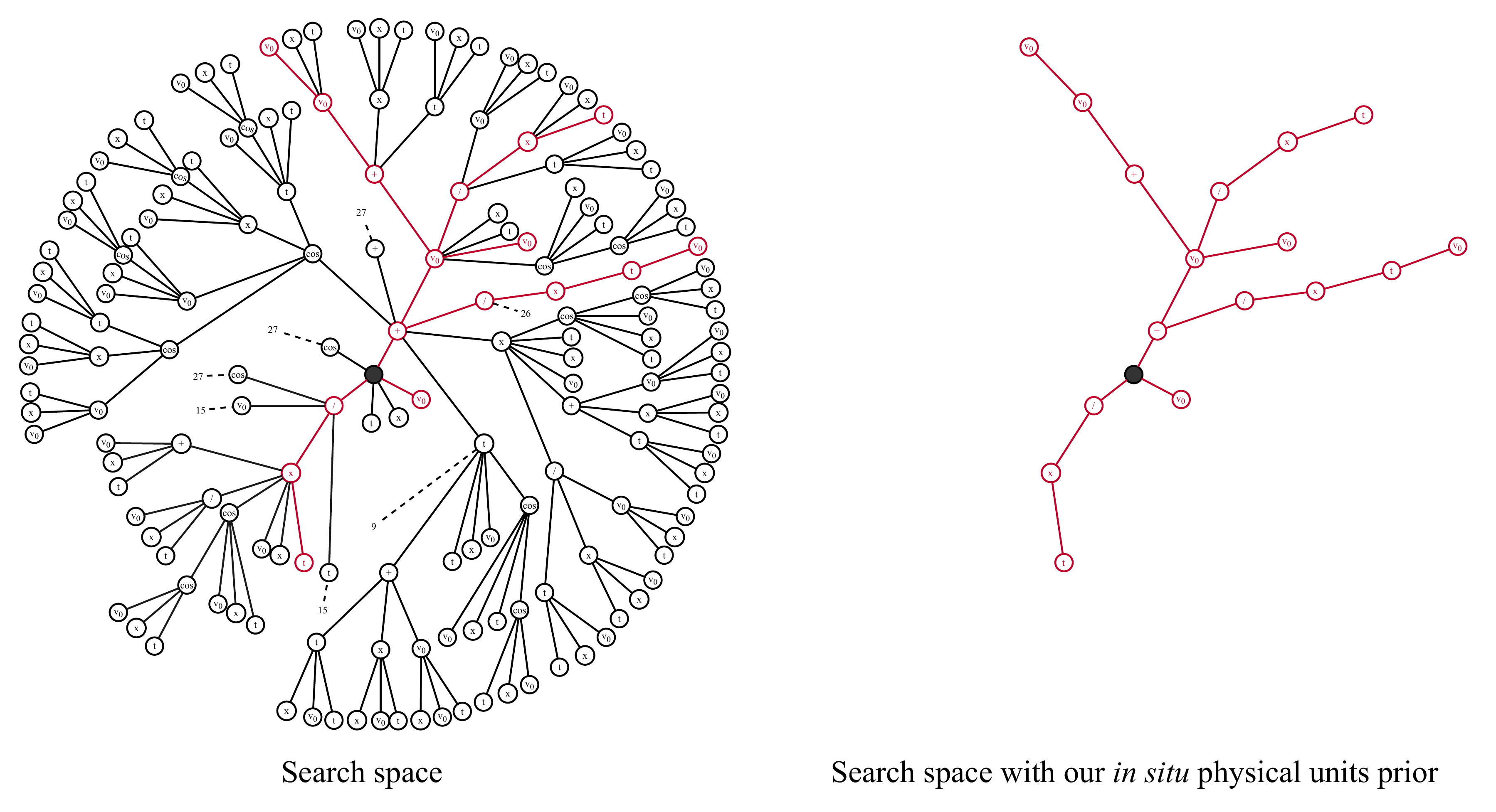}
\end{center}
\caption{Illustration of the symbolic expression search space reduction enabled by our \insitu physical units prior. We represent paths (in prefix notation) leading to expressions with physically-possible units (in red), a sample of the paths that lead to expressions with unphysical units (in black) with other unphysical paths redacted for readability summarized with dotted lines and their total number. Here we consider the recovery of a velocity $v$ using a library of symbols $\{ +, /, \cos, v_0, x, t \}$ where $v_0$ is a velocity, $x$ is a length, and $t$ is a time (limiting ourselves to 5 symbol long expressions for readability). This reduces the search space from 268 expressions to only 6.}
\label{fig:space_reduction_illustration}
\end{figure*}


Galileo famously intuited in Opere Il Saggiatore \citep{Galilei1623} that the book of the Universe ``\`e scritto in lingua matematica''. Ever since, it has been a central concern of physics to attempt to explain the properties of nature in mathematical terms, by proposing or deriving mathematical expressions that encapsulate our measurements from experiment and observation. This approach has proven to be immensely powerful. Through trial and error over the centuries, the great masters of physics have developed and bequeathed us a rich toolbox of techniques that have allowed us to understand the world and build our modern technological civilization. But now, thanks to the development of modern deep learning networks, there is hope that this endeavor could be accelerated, by making use of the fact that machines are able to survey a vastly larger space of trial solutions than an unaided human.


Of course, since the beginning of the computer revolution, many methods have been developed to fit coefficients of linear or non-linear functions to data (see, e.g., \citealt{NumericalRecipes}). While such approaches are undoubtedly very useful, the procedures we wish to discuss in the present contribution are more general, in the sense that they aim to find the functions themselves, as well as any necessary fitting coefficients. In particular, we wish to infer a free-form symbolic analytical function $f: \mathbb{R}^n \longrightarrow \mathbb{R}$ that fits $y = f(\mathbf{x})$ given $(\mathbf{x}, y)$ data. In computer science, these procedures are generally referred to as ``Symbolic Regression'' (SR).

\subsection{Motivations from physics and Big Data}
\label{subsec:motivations}

Although there are multiple demonstrations of the capabilities of SR in physics (e.g., \citealt{SR_phy_demo_TowardAIPhysicist, SR_phy_demo_ConservationLaws_AIFeynman, SR_phy_demo_NonConservativeNewPhysics, SR_phy_demo_OrbitalMechanics_PySR, SR_phy_demo_hamiltonian_DSRbased, Exhaustive_SR, SR_appli_fluid_real_data}) and astrophysics (e.g., \citealt{SR_appli_astro_scales, SR_appli_exoplanet_units, SR_appli_subhalo, SR_appli_galaxyhalo, SR_appli_haloHI, SR_appli_RAR_relation, SR_appli_graviwave}), to date, symbolic regression has never been used to discover new physical laws from astrophysical measurements. Yet this may change thanks to new observational missions and surveys such as Gaia \citep{GaiaMission}, Euclid \citep{EuclidMission}, LSST \citep{LSSTMission, LSSTsciencebook} and SKA \citep{SKAgoals}. With these and other large surveys, our field is entering a new era of data abundance, and there is considerable excitement at the possibility of identifying new empirical laws from these unprecedentedly rich and intricate datasets that could eventually lead to the discovery of new physics. However, the colossal amount of data also presents significant conceptual challenges. Although deep learning will allow us to extract valuable information from the large surveys, it is both blessed and plagued by the underlying neural networks that are one of its most potent components. Neural networks are flexible and powerful enough to model any physical system (that can be described as a Lebesgue integrable function \citealt{Lu2017}) and work in high dimensions, but they unfortunately largely consist of non-interpretable black boxes. Clearly, interpretability and intelligibility are of great importance in physics, which begs the question: how can one harness information from these large datasets while retaining their ability to interpret and connect with theory? After training a deep neural network to fit a dataset, can one open the black box, to understand the physics modelled inside?

\subsection{Symbolic regression}
\label{subsec:symbolic_regression}


Symbolic regression addresses these issues by producing compact, interpretable and generalizable models. Indeed, the goal is to find very simple prescriptions such as Newton's law of universal gravitation that can explain well a vast number of experiments and observations. There are many advantages to discovering physical laws in the form of succinct mathematical expressions rather than large numerical models:
\begin{itemize}

\item {Compactness: } SR methods can produce extremely compact models, \eg with expressions of containing $\sim 10^1$ symbols \citep{SRBench} which is on par with the typical length of expressions in the Feynman Lectures on Physics \citep{FeynmanLectures} for example which is of 16 (with the higher end of SR methods producing expressions well below a length of $10^3$). In contrast numerical models such as neural networks typically rely on many more parameters. This makes the models computationally inexpensive to run and in principle also enables SR to correctly recover the exact underlying mathematical expression of a dataset using much less data than traditional machine learning approaches \citep{SR_Generalization_SmallDataset} and with a robustness towards noise even for perfect model recovery \citep{SR_appli_fluid_real_data, SRBench}.
\item {Generalization:} In addition, unless the target equations consist of arbitrarily long polynomials, the compact expressions produced by SR are less prone to overfitting on measurement errors and are much more robust and reliable outside of the fitting range provided by the data than large numerical models, showing overall much better generalization capabilities as demonstrated in \citep{EQL_SymbolsInNN, Kamienny_EndToEndSR, Kamienny_SR_for_RL, SR_Generalization_SmallDataset} (we will provide an example of this in Section \ref{subsec:astro_results}). This makes SR a potentially powerful tool to discover the most concise and general representation of the measurements.
\item {Intelligibility \& interpretability:} Since the models produced by SR consist of mathematical expressions, their behavior is intelligible to us, unlike large numerical models. This is of enormous value in physics \citep{SR_phy_demo_TowardAIPhysicist} as SR models may enable one to connect newly discovered physical laws with theory and make subsequent theoretical developments. More broadly, this approach fits into the increasing push towards intelligible \citep{symbolic_repr_evaluation}, explainable \citep{Arrieta2020} and interpretable \citep{Murdoch2019} machine learning models, which is especially important in fields where such models can affect human lives \citep{UE2021,USA2022}.

\end{itemize}


However, although the prospect of using SR for discovering new physical laws may be very appealing, it is also extremely challenging to implement. It is useful to consider the difficulty of this problem if one were to approach it in a naive way. Suppose in the trial analytic expressions, we allow for an expression length of 35 symbols (as we will do below), and that there are 15 different variables or operations (e.g. $x$, $+$, $-$, $\times$, $/$, $\sin$, $\log$, ...) to chose from for each symbol (which is on par with what we will do below). A naive brute-force attempt to fit the dataset might then have to consider up to $15^{35} \approx 1.5\times10^{41}$ trial solutions which is obviously vastly beyond our computational means to test against the data at the present day or at any time in the foreseeable future, making SR an ``NP hard'' (nondeterministic polynomial time) problem \citep{SRisNPhard}. Furthermore, one has to account for the optimization of free constants in the proposed expressions. The obvious conclusion one draws from these considerations is that symbolic regression requires one to develop highly efficient strategies to prune poor guesses. 

\subsection{Physical Symbolic Regression}
\label{subsec:PhySO}

There are multiple approaches to SR (detailed in Section \ref{sec:related_works}) which are capable of generating accurate analytical models. However, in the context of physics, we have the additional requirement that our equations must be balanced in terms of their physical units, as otherwise the equation is simply non-sensical, irrespective of whether it gives a good fit to the numerical values of the data. Although powerful, to the best of our knowledge, all of the available SR approaches spend most of their time exploring a search space where the immense majority of candidate expressions are unphysical in terms of units and thus often end up producing unphysical models (with the exception of approaches in which variables are rendered dimensionless beforehand as discussed in sub-section \ref{subsec:units_prior}). A very simple solution to this problem would have been to use an existing SR code, and check \posthoc whether the proposed solutions obey that constraint. But not only does that constitute an immense waste of time and computing resources, which could render many interesting SR tasks impossible, it also makes a significant fraction of the resulting ``best'' analytical models unusable and uninterpretable. We note that for the sake of clarity, throughout this paper we refer to a system of \emph{unique} quantities such as physical dimensions $\{L, M, T, I, \Theta, N, J\}$ \ie with physical units $\{\text{m}, \text{kg}, \text{s}, \text{A}, \text{K}, \text{mol}, \text{cd}\}$\, a subset thereof, or problem-specific quantities such as $\{L, V, \rho, P, v\}$ \ie with physical units $\{\rm m, m^3, kg/m^{-3}, Pa, m.s^{-1}\}$ as ``units'' \footnote{Although this can also be extended to systems with non-physical quantities, such as $\{\text{scalar}, \text{vector}, \text{matrix}\}$ or even $\{\text{dollars}, \text{capita}, \text{annum}\}$.}. 

At first glance, one could think of the units constraints as severe restrictions that limit the capabilities of SR as they would prevent the generation of unphysical intermediary expressions. However, in this work we show that respecting physical constraints actually helps improve SR performance not only in terms of interpretability but also in accuracy by guiding the exploration of the space of solutions towards exact analytical laws. This is consistent with the studies of \citep{PetersenDSR, DSR_priors, Exhaustive_SR_wconstraints} who found that using \insitu constraints during analytical expression generation is much more efficient as it vastly reduces the search space of trial expressions
(though we note that incorporating such constraints in those frameworks would not be straightforward as one would need to recompute the whole relational graph representing an analytical expression and its underlying units constraints each time a new symbol is added).

Here we present our Physical Symbolic Optimization framework (\PhySO) which was designed from the beginning to incorporate and take full advantage of physical units information during symbolic regression by storing and managing information related to dimensional analysis. This addresses in part the combinatorial challenge discussed above in sub-section \ref{subsec:symbolic_regression}. Our \PhySO\ framework includes the units constraints \insitu during the equation generation process, such that only equations with balanced units are proposed {\it by construction}, thus also greatly reducing the search space as illustrated in Figure \ref{fig:space_reduction_illustration}.

Although our framework could be applied to virtually any one of the SR approaches described in Section \ref{sec:related_works}, we chose to implement our algorithm in \texttt{PyTorch} (\citealt{pytorch}, currently the most popular deep learning library in research, \citealt{pytorch_popularity}) building our method from scratch yet using some of the mathematical principles and key strategies pioneered in the state-of-the-art Deep Symbolic Regression framework proposed in \citet{PetersenDSR} and \citet{SR_PG_improvements} which rely on reinforcement learning via a risk-seeking policy gradient (which is based on \citealt{risk_averse_RL}).

In the present study, we develop a foundational symbolic embedding for physics that enables the entire expression tree graph to be tackled, as well as local units constraints. Unlike previous attempts to consider units in which datasets were rendered dimensionless before applying standard SR techniques \citep{AIFeynman, SR_appli_exoplanet_units, SciMED}, our approach allows us to anticipate the required units for the subsequent symbol to be generated in a partially composed mathematical expression. By adopting this approach, we not only focus on training a neural network to generate increasingly precise expressions, as in \citet{PetersenDSR}, but we also generate labels of the necessary units and actively train our neural network to adhere to such constraints. In essence, our method equips the neural network with the ability to learn to select the appropriate symbol in line with local units constraints.

To the best of our knowledge such a framework was never built before. This constitutes a first step in our planned research program of building a powerful general-purpose symbolic regression algorithm for astrophysics and other physical sciences. Our aim here is to present the algorithm to the community, show its workings and its potential, while leaving concrete astrophysical research applications to future studies. 

The layout of this study is as follows. We first provide a brief overview of the recent SR literature in Section~\ref{sec:related_works}. Our \PhySO\ framework is described in detail in Section~\ref{sec:method}, in Section~\ref{sec:feynman} we apply it to a benchmark of 120 equations from the Feynman Lectures on Physics and compare it to 17 other popular SR algorithms, reporting state-of-the-art performance. In Section~\ref{sec:case_studies} we showcase \PhySO's capabilities on a panel of astrophysical test cases and perform an ablation study. Finally in Sections~\ref{sec:discussion} and ~\ref{sec:conclusions} we discuss the results and draw our conclusions.


\section{Related works -- a brief survey of modern Symbolic Regression}
\label{sec:related_works}


SR has traditionally been tackled using genetic programming where a population of candidate mathematical expressions are iteratively improved through operations inspired by natural evolution such as natural selection, crossover, and mutation. This type of approach includes the well known Eureqa software \citep{EureqaPaper2009, EureqaPaper2011_AFP} (see \citealt{EureqaAstroBenchmark} for a benchmark of Eureqa's capabilities on astrophysical test cases), as well as more recent works \citep{PySR, ITEA, EPLEX, FEAT, SBP_GP, Cranmer2020_GNN, GP_GOMEA, gplearn, OPERON}. In addition, SR has been implemented using various methods ranging from brute force to (un-)guided Monte-Carlo, all the way to probabilistic searches \citep{FFX_SR, 
Exhaustive_SR_wconstraints, Exhaustive_SR, Grammar_prior_wMC_SR, BSR_Bayesian_SR}, as well as through problem simplification algorithms \citep{Divide_and_conquer, GSR}. 

Given the great successes of deep learning techniques in many other fields, it is not surprising that they have now been applied to symbolic regression, and now challenge the reign of Eureqa-type approaches \citep{SRBench, Benchmark_SR_phy}. Multiple methods for incorporating neural networks into SR have been developed, ranging from powerful problem simplification schemes \citep{AIFeynman, AIFeynman2, Cranmer2020_GNN}, to end-to-end symbolic regression methods where a neural network is trained in a supervised manner to map the relationship between datasets and their corresponding symbolic functions \citep{Kamienny_EndToEndSR, Biggio_EndToEndSR, NeSymReS_EndToEndSR, SymFormer_EndToEndSR, Meta_EndToEndSR_MathsRecurrent, Kamienny_learning_mutations, Kamienny_learning_wassumptions, Becker_PDE_SR_EndToEnd, FeynmanDiagram_to_Seq, Toyota_EndToEndSR}, all the way to incorporating symbols into neural networks and sparsely fitting them to enable interpretability or to recover a mathematical expression \citep{Martius_SymbolsInNN, Sindy_SymbolsInNN, Zheng_SymbolsInNN_DSRbased, EQL_SymbolsInNN, Valle_SymbolsInNN, Kim_SymbolsInNN, Panju_SymbolsInNN, SISSO}. See \citep{SRBench, SR_interpretability_review, SR_review_Angelis}, for recent reviews of symbolic regression algorithms.

While some of the aforementioned algorithms excel at generating very accurate symbolic approximations, the reinforcement learning based deep symbolic regression framework proposed in \citet{PetersenDSR} is the new standard for exact symbolic function recovery, particularly in the presence of noise \citep{SRBench, Benchmark_SR_phy}. This has resulted in a number of studies in the literature built on this framework \citealt{SR_PG_improvements, SR_of_RL_policies_DSRbased, DSR_wikipedia_prior, DSR_priors, uDSR_DSRbased, Du_PDE_SR_DSRbased, SR_phy_demo_hamiltonian_DSRbased, Zheng_SymbolsInNN_DSRbased,
SR_of_RL_policies_DSRbased, SR_of_controller_DSRbased}.


\section{Method}
\label{sec:method}

\begin{figure*}
\begin{center}
\includegraphics[angle=0, clip, width=\hsize]{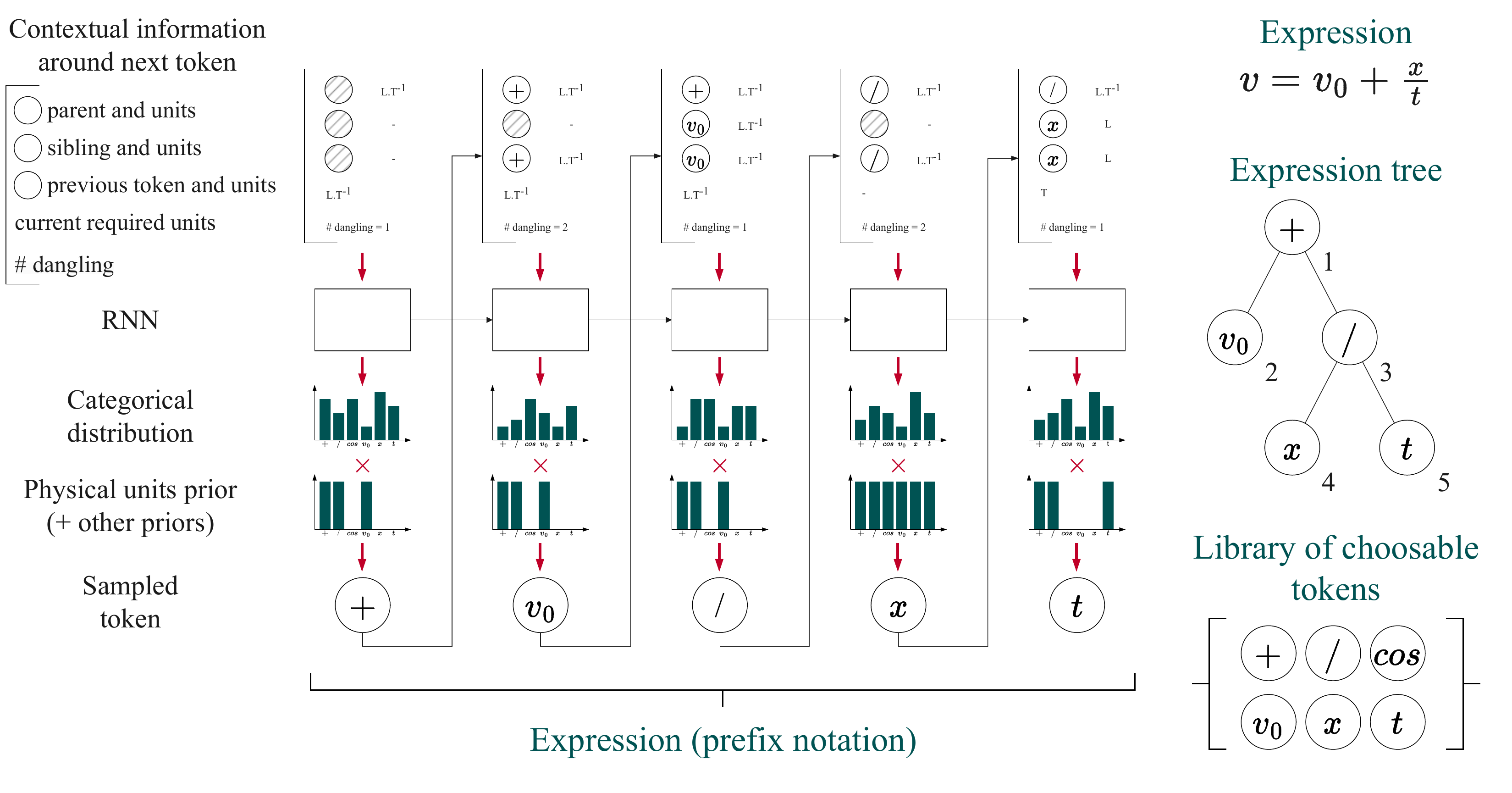}
\end{center}
\caption{Expression generation sketch. The process starts at the top left RNN block. For each token, the RNN is given the contextual information regarding the surroundings of the next token to generate, namely: the parent, sibling and previously sampled token along with their units, the required units for the token to be generated and the dangling number (\ie the minimum number of tokens needed to obtain a valid expression). Based on this information, the RNN produces a categorical distribution over the library of available tokens (top histograms) as well as a state which is transmitted to the RNN on its next call. The generated distribution is then masked based on local units constraints (bottom histograms), forbidding tokens that would lead to nonsensical expressions. The resulting token is sampled from this distribution, leading to the token `$+$' in this example. Repeating this process, from left to right, allows one to generate a complete physical expression, here $[+, v_0, /, x, t]$ which translates into $v_0+x/t$ in the infix notation we are more familiar with.}
\label{fig:embedding}
\end{figure*}

Considering the success of deep reinforcement learning methods in accurately recovering exact symbolic expressions, which is particularly important in the field of physics where precise physical law recovery is crucial, we have chosen to incorporate this methodology into the machine learning component of our physical symbolic regression approach. In sub-section \ref{subsec:embedding}, we describe how we generate analytical expressions from a recurrent neural network (RNN). Sub-section \ref{subsec:units_prior} provides details about the algorithm we use to generate \insitu units constraints, which are used to teach the RNN dimensional analysis rules and help to reduce the search space. In sub-section \ref{subsec:learning}, we describe the reinforcement learning strategy we adopted to make our RNN not only produce accurate expressions but also physically meaningful ones. Finally, we give computational details regarding our Physical Symbolic Optimization implementation (\texttt{PhySO}) in sub-section \ref{subsec:computational}.

\subsection{Generating symbolic expressions}
\label{subsec:embedding}

Symbolic expressions can be regarded as binary trees where each node represents a symbol of the expression in the library of available symbols, i.e., an input variable (\eg $x$, $t$), a constant (\eg $v_0$) or an operation (\eg $+$, $-$, $\times$, $/$, $\sin$, $\log$, ...). In this representation, input variables and constants can be referred to as terminal nodes or symbols (having no child node), operations taking a single argument (\eg $\sin$, $\log$, ...) are unary symbols (having one child node) and operations taking two arguments (\eg $+$, $-$, $\times$, $/$, ...) are binary symbols. By considering each node first in depth and then left to right, one can compute a one dimensional list \ie a prefix\footnote{This is also called ``Polish'' notation, and can be converted to a tree representation or the ``infix'' notation which we are more familiar with, as there is a one-to-one relationship between them.} notation in which operators are placed before the corresponding operands in the expression, alleviating the need for parentheses. Using the prefix notation and treating symbols, referred to as tokens, as categories allows us to treat any expression as a mere sequence of categorical vectors. E.g., considering short toy library of tokens $\{+, \cos, x\}$, the operator $+$ can be encoded as $[1, 0, 0]$, the function $\cos$ as $[0, 1, 0]$ and the variable $x$ as $[0, 0, 1]$.

As in previous deep symbolic regression studies (\eg \citep{Kamienny_EndToEndSR, SymFormer_EndToEndSR, PetersenDSR, Du_PDE_SR_DSRbased, SR_phy_demo_hamiltonian_DSRbased}, treating mathematical expressions as sequences allows us to employ traditional natural language processing techniques to sample them. Token sequences are generated by using an RNN, which in essence, is a neural network that can be invoked multiple times to create a logical chain of similar operations. At each invocation $i<N$ ($N$ representing the maximum number of steps), the RNN generates a time-dependent output and a corresponding memory state $S_i$. The RNN takes as input some time dependent observations $O_i$ \footnote{We refer to `observations' in the context of reinforcement learning, here pertaining to contextual analytical information related to the expression being generated, rather than to the scientific data being fitted.} as well as the state of the previous call $S_{i-1}$. In practice, we use the RNN to generate a categorical probability distribution over the library of available tokens, which we then simply sample to draw a definite token. Once a token is generated, we feed the minimum number of tokens still needed to obtain a valid analytical expression (\ie the number of dangling nodes), the token's properties and the properties of its surroundings as observations for the next RNN call. Namely, we give the nature of the token which was sampled at the previous step (since the RNN does not have access to this information which is derived from a stochastic process)\footnote{Not providing this information typically hinders performance.}, the sibling (if any at this step) and parent tokens of the token to be generated in a tree representation, to which in the context of our \PhySO\ framework we add the physical units of all of these tokens and the units required for the token to be generated so as to respect the units rules. This allows the inner mechanisms of the neural network to take into account not only the local structure of the expression for generating the next token, but also to take into account the local units constraints. The process described above can be repeated multiple times until a whole token function is generated in prefix notation, as illustrated in Figure \ref{fig:embedding}.

It is important to note here that one can artificially tune the generated categorical distribution to incorporate prior knowledge \insitu while expressions are being generated. One can for example zero-out the probability of some token depending on the context encoded in the expression tree being generated, thus greatly reducing search space \citep{PetersenDSR, DSR_priors}. We therefore adopt priors that force expressions sizes to be $< 35$ tokens long, encourage expressions to be ``concise'' through a soft length prior consisting of a Gaussian of variance $\sigma^2=5$ centered around a length of 8, to contain no more than 2 levels of nested trigonometric operations (\eg forbidding $\cos(f.t + \sin(x/x0 + \tan(\placeholder)))$ but still allowing $\cos(f.t + \sin(x/x0))$), contain no self nesting of exponent and log operators (\eg forbidding $e^{e^\placeholder}$) and forbid useless inverse unary operations (\eg forbidding $e^{\log \placeholder}$). It is worth noting that the combination of priors we employ can conflict in some cases, in which case we discard the resulting candidate (\eg the physical units prior detailed below could require a certain number of tokens to satisfy the units constraints, which could conflict with the length prior requiring the expression to be terminated prematurely).

In addition to the above priors whose formulation depends on the local tree structure (parent, sibling, ancestors), our method is able to accommodate any priors that take into account the entire tree structure without having to recompute it from scratch at each step. This is rendered possible by the fact that contrary to other deep learning based SR algorithms, in the \PhySO\ framework we compute and keep track of the full graph of the tree representation and its underlying grammatical information (such as units, symbol types like functions, free parameters, fixed constants or the number of arguments a symbol requires) while the expression is being generated, as it is an essential ingredient to compute units constraints as detailed in the following sub-section. Note that this also enables \PhySO\ to accommodate any future prior relying on such information.

\subsection{\Insitu physical units constraints}
\label{subsec:units_prior}

\def\parent{{\tau_p}}
\def\sibling{{\tau_s}}
\def\self{{\tau_i}}

\def\uparent{{\Phi_p}}
\def\usibling{{\Phi_s}}
\def\uself{{\Phi_i}}

\algnewcommand{\AND}{\textbf{ and }}
\algnewcommand{\NOT}{\textbf{ not }}
\algnewcommand{\OR}{\textbf{ or }}

\begin{algorithm}
\DontPrintSemicolon

\KwIn{(In)-complete expression $\{\tau_j\}_{j<N}$, Position of token $i$}
\KwOut{Required physical units $\uself$ of token at $i$}

\SetKwBlock{Begin}{Function}{end function}
\Begin($\text{ComputeRequiredUnits} {(} \{\tau_j\}_{j<N}, i {)}$)

{
$p \gets \text{PositionOfParent} (i)$ \;
$s \gets \text{PositionOfSibling} (i)$ \;
$\uparent \gets \text{Units} (\parent)$ \;
$\usibling \gets \text{Units} (\sibling)$ \;
$\text{NodeRank} \gets \text{1 if left side node and 2 if right node}$ \;
$\text{AdditiveTokens} \gets \{+, -\}$ \;
$\text{MultiplicativeTokens} \gets \{\times, /\}$ \;
$\text{PowerTokens} \gets \{1/\placeholder, \sqrt{\placeholder}, \placeholder^n\}$ \;
$\text{PowerValues} \gets \{1/\placeholder : -1, \sqrt{\placeholder} : 1/2, \placeholder^n : n\}$ \;
$\text{DimensionlessTokens} \gets \{\cos, \sin, \tan, \exp, \log\}$ \;

\uIf {\textup{$\parent$ is in AdditiveTokens \AND $\usibling$ is known}}
    {
    $\uself \gets \usibling$\;
    }

\uElseIf{\textup{$\parent$ is in AdditiveTokens \AND $\uparent$ is free \AND NodeRank is $2$ \AND $\usibling$ is free}}
    {
    BottomUpUnitsAssignement(start = $s$, end = $i-1$)\;
    $\uself \gets \usibling$\;
    }
\uElseIf{\textup{$\uparent$ is free \AND $\parent$ is \NOT in MultiplicativeTokens \AND $\sibling$ is \NOT a placeholder}}
    {
    $\uself \gets \text{free}$;
    }

\uElseIf{\textup{$i = 0$}}
    {
    $\uself \gets \text{Units(root)}$\;
    }

\uElseIf{\textup{$\parent$ is in AdditiveTokens}}
    {
    $\uself \gets \uparent$\;
    }

\uElseIf{\textup{$\parent$ is in PowerTokens}}
    {
    $n \gets \text{PowerValues}[\parent]$\;
    $\uself \gets \uparent / n$\;
    }

\uElseIf{\textup{$\uparent = \mathbf{0}$ \OR $\parent$ is in DimensionlessTokens}}
    {
    $\uself \gets \mathbf{0}$\;
    }

\uElseIf{\textup{$\parent$ is in MultiplicativeTokens}}
    {
    \uIf{\textup{$\self$ is a placeholder \AND $\sibling$ is a placeholder}}
        {
        $\uself \gets \text{free}$\;
        }
    \uElseIf{\textup{NodeRank is $1$}}
        {
        $\uself \gets \text{free}$\;
        }
    \uElseIf{\textup{$\uparent$ is free}}
        {
        $\uself \gets \text{free}$\;
        }
    \Else
        {
        BottomUpUnitsAssignement(start = $s$, end = $i-1$)\;
        \uIf{\textup{$\self$ is $\{ \times \}$}}
            {
            $\uself \gets \uparent - \usibling$;
            }
        \ElseIf{\textup{$\self$ is $\{ / \}$}}
            {
            $\uself \gets \usibling - \uparent$;
            }
        }
    }

\Return{$\uself$}
}
\caption{\Insitu units requirements algorithm}
\label{alg:algo_units}
\end{algorithm}
\newcommand{\ope}{\textcolor{black}{\text{op}_\mathbf{0}}}
\newcommand{\varA}{{\tau_A}}
\newcommand{\varB}{{\tau_B}}
\newcommand{\vary}{{y}}
\newcommand{\uA}{{\Phi_A}}
\newcommand{\uB}{{\Phi_B}}
\newcommand{\uy}{{\Phi_y}}

\begin{table}
\begin{center}
\begin{tabular}{cc}
\multicolumn{2}{c}{Dimensional analysis rules} \\ \hline
Expression           & Units               \\ \hline
$\varA \pm \varB$    & $\uA$ or  $\uB$     \\
$-\varA$             & $\uA$               \\
$\varA \times \varB$ & $\uA + \uB$         \\
$\varA / \varB$      & $\uA - \uB$         \\
$\varA^n$            & $n \times \uA$      \\
$\ope$$(\varA)$      & $\mathbf{0}$        \\ \hline
                     &                     \\
                     &                     \\
\multicolumn{2}{c}{Units requirements rules} \\ \hline
Expression               & Requirement       \\ \hline
$\varA \pm \varB$        & $\uA = \uB$        \\
$\vary = \varA$          & $\uy = \uA$        \\
$\ope$$(\varA)$          & $\uA = \mathbf{0}$ \\ \hline
\end{tabular}
\end{center}
\caption{Dimensional analysis prescriptions to enforce. With $\varA$, $\varB$, $y$, $\uA$, $\uB$, $\uy$ referring to two nodes, the output variable and the powers of their units vectors, $\ope$ denoting a dimensionless operation (\eg $\{\cos,\sin, \exp, \log \}$) and $\varA^n$ representing any power operation (including \eg $1/\varA = \varA^{-1}$, $\sqrt{\varA} = \varA^{\frac{1}{2}}$)}
\label{table:dimensional_analysis}
\end{table}

Our work is part of the broader field of grammar-guided SR \citep{Muhammad_grammar_SR, Grammar_prior_wMC_SR, RL_grammar_SR, Korns_grammar_SR, Hoai_grammar_SR, Manrique_grammar_SR, Worm_grammar_SR} which aims at constraining the symbolic arrangement of mathematical expressions based on domain specific rules. Specifically and as discussed above, in physics we already know that some combinations of tokens are not possible due to units constraints. For example, if the algorithm is in the process of generating an expression in which a velocity ($v_0$) is summed with a length ($x$) divided by a token or sub-expression which is still to be generated ($\placeholder$):
\begin{equation}
    v_0 + \frac{x}{\placeholder} \, ,
\end{equation}
then based on the expression tree (as shown in Figure \ref{fig:embedding}), we already know that that $\placeholder$ must be a time variable or a more complicated sub-tree that eventually ends up having units of time, but that it is definitely not a length or a dimensionless operator such as the $\log$ function.

Computing such constraints \insitu \ie in incomplete, only partially sampled trees (containing empty placeholder nodes) is much harder than simply checking \posthoc if the units of a given equation make sense, because in some situations it is impossible to compute such constraints until later on in the sequence, leaving the units of some nodes {\it free} (\ie compatible with any units at this point in the sequence). For example, it is impossible to compute the units requirement in the left child node of a (binary) multiplication operator token $\placeholder\times\triangle$, as any units in the $\placeholder$ left child node could be compensated by units in the $\triangle$ right child node. Following the dimensional analysis rules summarized in Table \ref{table:dimensional_analysis}, we devised Algorithm \ref{alg:algo_units}. This algorithm gives the pseudo-code of the procedure we devised to compute the required units whenever possible and leaving them as {\it free} otherwise. The procedure is applied to a token at position $i<N$ in an incomplete or complete sequence of tokens $\{\tau_j\}_{j<N}$ of size $N$, knowing the units of terminal nodes and of the root node (\eg respectively $\{v_0, x, t\}$ and $\{v\}$ in the example of Figure \ref{fig:embedding}). The sequence may be partially made up of placeholder tokens of yet undetermined nature (representing dangling nodes). Running algorithm \ref{alg:algo_units} before each token generation step allows one to have a maximally informed expression tree graph in terms of units.

Having access \insitu to the (required) physical units of tokens allows us not only to inform the neural network of our expectations in terms of units as well as to feed it units of surrounding tokens, thus allowing the model to leverage such information, but also to express a prior distribution over the library. This enables the algorithm to zero-out the probability of forbidden symbols that would result in expressions that violate units rules. Combining this prior distribution with the categorical distribution given by the RNN while expressions are being generated results in a system where {\it by construction} only correct expressions with correct physical units can be formulated and learned on by the neural network.\\

We acknowledge a previous attempt by \citep{AIFeynman} in the \texttt{AI Feynman} algorithm to consider units in the context of SR. The approach adopted by \texttt{AI Feynman} addresses symbolic regression problems by first transforming the variables to make them dimensionless, often leading to a reduction in the number of variables and allowing the generation of physically balanced expressions. However, if this method fails, the algorithm reverts to the original problem setup. This results in \texttt{AI Feynman} resorting to fitting high-order polynomials or complicated expressions that although very accurate lack physical meaning from a dimensional analysis perspective most of the time when it is not able to find a perfect fit solution. For instance, even in the shorter range of expressions it proposes, one can find equations such as $K = \arcsin(0.169*e^{-3.142*m + w})$ where $K$, $m$ and $w$ denote an energy, a mass and a velocity for Feynman problem I.13.4 (details about the Feynman symbolic regression problems can be found in sub-section \ref{subsec:feynman_protocol}). In contrast, \PhySO\ is designed to yield only physically plausible expressions by construction all of the time. Contrary to \texttt{AI Feynman}, \PhySO\ works on dimensional data by leveraging constraints on the functional forms while generating expressions as outlined in Table \ref{table:dimensional_analysis}. It is worth noting, however, that making problems dimensionless, as implemented in \texttt{AI Feynman}, is a valuable approach that can work in pair with any symbolic regression method to ensure outputs are not non-sensical.\\

Indeed, it could be argued that we could have tackled the physical units validity of expressions in SR by taking advantage of the Buckingham $\Pi$ theorem \citep{Buckingham_pi_theorem}, with variables and constants rendered dimensionless by means of multiplicative operations amongst them. Such an approach can actually be adopted as a preliminary step in conjunction with any SR framework (see, \eg\ \citealt{SR_appli_exoplanet_units, SciMED}). However, although working with so called $\Pi$ groups ensures the generation of physically valid expressions (since all terms become dimensionless), it simultaneously removes constraints imposed by dimensional analysis, complicating the SR process. It is interesting to note that nature (or at least physics) is not dimensionless, so information is lost during the process of making variables and constants dimensionless, preventing us from leveraging the powerful constraints on the functional form associated with this dimensional information.
Drawing from the example presented in \citet{AIFeynman}, let us consider a dataset associated with the target expression
\begin{equation}
F = \frac{G m_1 m_2}{(x_2 - x_1)^2 + (y_2 - y_1)^2 + (z_2 - z_1)^2} \, .
\end{equation}
When rendered dimensionless, the target expression becomes 
\begin{equation}
y = \frac{\frac{m_2}{m_1}}{(\frac{x_2}{x_1}-1)^2 + (\frac{y_2}{x_1}-\frac{y_1}{x_1})^2+ (\frac{z_2}{x_1}-\frac{z_1}{x_1})^2} \, .
\end{equation}
While this transformation decreases the number of input variables to $\{\frac{m_2}{m_1}, \frac{x_2}{x_1}, \frac{y_2}{x_1}, \frac{z_1}{x_1}\}$, it simultaneously nullifies the inherent dimensional analysis constraints. Consequently, the SR algorithm could potentially produce expressions such as \(\frac{m_2}{m_1} - \frac{x_2}{x_1}\) or \((\frac{x_2}{x_1} -1 )^2 + \frac{y_2}{x_1}\). In contrast, with our \insitu constraints, lengths could only be summed with lengths terms, similarly, squared lengths could only be summed with squared lengths and having \(G m_1 m_2\) in the numerator would be enforced by the requirement of the expression being homogeneous to a force. In essence, while rendering variables dimensionless ensures physicality of the expressions, it simultaneously relinquishes valuable constraints on their functional forms.\\

Finally, we note that after the first submission of our paper, two approaches similar to ours were presented, the first working in pair with a sparsity fitting method \citep{SISSO_DA} and the second working in pair with a probabilistic search method \citep{DA_grammar_SR}.


\subsection{Learning}
\label{subsec:learning}

\def\HyperparamRiskFactor{{$5$\ }} 
\def\HyperparamNoRiskFactor{{$95$\ }} 

One might imagine that symbolic regression problems could be solved by directly optimizing the choice of symbols to fit the problem, using the auto-differentiation capabilities of modern machine learning frameworks\footnote{Most machine learning tasks use the differentiability of the implemented model with respect to the data to implement a (stochastic) gradient descent towards an optimal model solution that fits the data best.}. Unfortunately this approach cannot be used for symbolic regression because the cost function is non differentiable (the choice of selecting say the $\sin$ function over $\log$ is not differentiable with respect to the data), which prevents one from using gradient descent. A practical solution is to use a neural network as a ``middle man'' to generate a categorical distribution from which we can sample symbols. One can then optimize the parameters of this neural network whose task is to generate these symbols according to fit quality and physical units constraints.

The training of the network that generates the distribution of symbols relies on the ``reinforcement learning'' strategy \citep{RL_book}, which is a common method used to train artificial intelligence agents to navigate virtual worlds such as video games\footnote{See, e.g., \texttt{https://www.youtube.com/watch?v=QilHGSYbjDQ}}, or master open-ended tasks \citep{RL_human_time_deepmind}. In the present context, the idea is to generate a set (usually called a ``batch'' in machine learning) of trial symbolic functions, and compute a scalar reward for each function by confronting it to the data. We can then require the neural network to generate a new batch of trial functions, encouraging it to produce better results by reinforcing behavior associated with high reward values, approximating gradients via a so-called ``policy'' (i.e., a quantitative strategy). The hope is that, by trial and error, the learnable parameters of the network will converge to values that are able to generate a symbolic function that fits the data well.

Following the insight by \cite{PetersenDSR}, we adopt the risk-seeking policy gradient along with the entropy regularization scheme found by \cite{SR_PG_improvements}. In essence, we only reinforce the best \HyperparamRiskFactor\% of candidate solutions, not penalizing the neural network for proposing the \HyperparamNoRiskFactor\% of other candidates, therefore maximizing the reward of the few best performing candidates rather than the average reward. With our chosen batch size of 10k, detailed in Table \ref{table:hyperparams}, this strategy reinforces the leading 500 candidates. This enables an efficient exploration of the search space at the expense of average performance, which is of particular interest in SR as we are often mostly concerned in finding the very best candidates in particular if the goal is exact symbolic recovery and do not care if the neural network performs well on average\footnote{This is contrary to many other applications of reinforcement learning (\eg robotic automation, video games) which can even sometimes require risk-adverse gradient policies (\eg self driving cars) \citep{risk_averse_RL}.}. This novel risk-seeking policy, inspired by \citep{risk_averse_RL} and first proposed by \cite{PetersenDSR}, has significantly boosted performance in symbolic regression.

It is worth noting that our approach reinforces candidates which are sampled based on not only the output of the RNN, but also the local units constraints derived from the units prior distribution, which ensures the physical correctness of token choices. As a result, our approach effectively trains the RNN to make appropriate symbolic choices in accordance with local units constraints, in a quasi supervised learning manner. This combined with the general reinforcement learning paradigm enables us to produce both accurate and physically relevant symbolic expressions.

We allow the candidate functions $f$ to also contain ``constants'' with fixed physical units specified by the user, but with free numerical values. These free constants allow us the possibility to model situations where the problem has some unknown physical scales. A (somewhat contrived) example from galactic dynamics could be if we were provided a set of potential values $\Phi$, and cylindrical coordinate values $(R,z)$ of some mystery function that was actually a simple logarithmic potential model:
\begin{equation}
    \Phi=\frac{1}{2}v_0^2 \ln \bigg(R_c^2+R^2+\frac{z^2}{q^2} \bigg) \, ,
\end{equation}
whose parameters are the velocity parameter $v_0$, the core radius $R_c$ and the potential flattening $q$. Of course, we will generally not know in advance either the number of such parameters that the correct solution requires, or their numerical values. Yet to be able to evaluate the loss of the trial functions $f$, we need to assign values to all such free ``constants'' they may contain. We accomplish this task by processing each trial function, with the L-BFGS \citep{LBFGS} optimization routine in {\it pytorch} (optimizing over 20 steps and using an mean squared error metric), leveraging the fact that we can encode the symbols of $f$ using {\it pytorch} functions. Since {\it pytorch} has in-built auto-differentiation, finding the optimal value of the constants via gradient descent is extremely efficient. 

Then, as in \citet{PetersenDSR} for each candidate $f$, we compute a reward $r$ that is representative of fit quality: $r = 1/(1+\text{NRMSE})$ where $\text{NRMSE}$ is the root mean squared error normalized by the deviation of the target $\sigma_y$: $\text{NRMSE} = \frac{1}{\sigma_y}\sqrt{\frac{1}{N}\sum_{i=1}^N (y_i - f(\mathbf{x}_i))^2}$. We apply the policy gradients by means of an Adam optimizer \cite{adam_optimizer} and use a long-short term memory (LSTM) type RNN \citep{LSTM}. Our additional learning hyper-parameters can be found in Table \ref{table:hyperparams}. It is worth noting that the empirically tuned batch size we found ($10$k) is larger than the one found by \cite{PetersenDSR} which was of $1$k. We attribute this to the very strong constraints offered by our \PhySO\ setup which require a strong exploration counterpart to avoid getting stuck in local minima. This helps ensure that the model does not prematurely converge by continuously reinforcing a locally optimal expression, but rather seeks more solutions until identifying the most favorable one. 

It is also worth noting that in the reinforcement learning framework, the the reward function can be considered as as a black box, which does not have to be differentiable, therefore one could use anything as the reward. For example, we can also include the complexity of the symbolic function in the reward function, so as to have a criterion akin to Occam's razor. But actually one could in principle implement many ideas into the reward function: symmetries, constraints on primitives or derivatives, fitness in a differential equation, the results of some symbolic computation using external packages such as \texttt{Mathematica} \citep{mathematica} or \texttt{SymPy} \citep{sympy}, behavior of the function when implemented an n-body simulation, and so on. Note that in the context of this work, although there are more sophisticated schemes to define complexity (see \eg \citealt{SR_complexity_metric}) we simply define it as length \ie the number of tokens appearing in the expression excluding parentheses or the number of nodes in a tree representation.

\begin{table}[]
\centering
    \begin{tabular}{cc}
    \hline
    \multicolumn{2}{c}{Learning parameters}                \\ \hline
    Batch size                & 10 000                     \\
    Learning rate             & 0.0025                     \\
    Entropy coefficient       & 0.005                      \\
    Risk factor               & \HyperparamRiskFactor \%   \\ \hline
    \end{tabular}
    \caption{Learning parameters.}
\label{table:hyperparams}
\end{table}

\subsection{Computational details}
\label{subsec:computational}

\begin{figure}
\begin{center}
\includegraphics[angle=0, clip, width=1.\hsize]{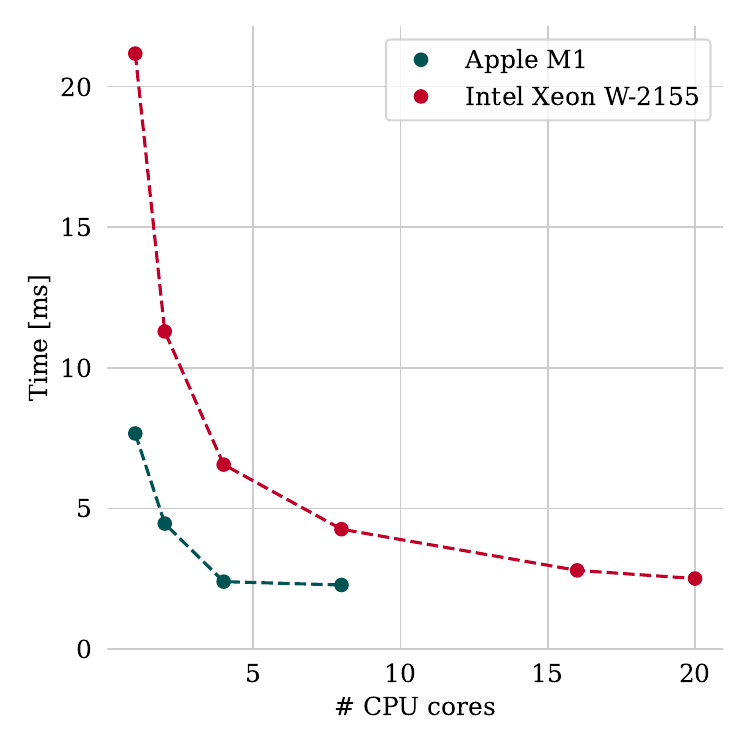}
\end{center}
\caption{Computational time optimizing free constants $\{a, b \}$ in $y = a \sin (b.x) + e^{-x}$ over 20 iterations using $10^3$ data points when running this task $10\ 000$ times in parallel with our \texttt{PhySO} algorithm on an Apple M1 laptop (a machine with 4 fast CPU cores) and an Intel Xeon W-2155 CPU (a machine with a high number of cores).}
\label{fig:physo_parallel_efficiency}
\end{figure}

Due to the number of trial expressions to evaluate at each iteration and considering that each expression must be evaluated multiple times to optimize its free constants, the optimization step is one of the main performance bottlenecks of our \texttt{PhySO}\ algorithm. This step was therefore parallelized across the batch, resulting in a free constants optimization time of a given expression typically being of the order of the $\ms$. We show an efficiency plot in a realistic scenario in Figure \ref{fig:physo_parallel_efficiency}. \footnote{The Apple M1 machine employed for the tests contains 4 high performance cores and 4 energy efficient cores, explaining the observed stagnation when increasing the cores count to 8.}

In addition, the management of symbolic information that is necessary to compute priors and contextual information to be passed to the neural network can also occupy a non-negligible part of the computational time. In \texttt{PhySO}, these operations are therefore vectorized across both equation lengths and batch.

Lastly, it is worth mentioning that upon concluding the exploration of the equation space, \texttt{PhySO} saves Pareto front equations (optimal solutions balancing fitness and low complexity) including the overall best fitting equation, the best-fitting equation across iterations, and stores a comprehensive log of all equations generated during the run.

\section{Feynman Benchmark}
\label{sec:feynman}

\definecolor{Tviolet}{HTML}{800080}
\definecolor{Tblue}{HTML}{3333ff}
\definecolor{Tbrown}{HTML}{843c0b}
\definecolor{Tred}{HTML}{cc0000}
\definecolor{Tgreen}{HTML}{006600}
\definecolor{Tgrey}{HTML}{000000}

\def\RL{{\textcolor{Tviolet}{RL}}}
\def\GP{{\textcolor{Tblue}{GP}}}
\def\Simp{{\textcolor{Tred}{Simp.}}}
\def\Sup{{\textcolor{Tbrown}{Sup.}}}
\def\DA{{\textcolor{Tgreen}{DA}}}
\def\MCMC{{\textcolor{Tgrey}{MCMC}}}
\def\Rand{{\textcolor{Tgrey}{Rand.}}}
\def\NeuroSym{{\textcolor{Tgrey}{NeuroSym}}}

\definecolor{lightgray}{gray}{0.9}

\begin{table*}
\begin{center}
\begin{tabular}{llll}
\toprule
                Method  &          Technique(s) &                                          Description &                   Reference  \\
\midrule
        \texttt{PhySO}  &             \RL, \DA  &                       Physical Symbolic Optimization &                        This work  \\
         \texttt{uDSR}  & \RL, \GP, \Simp, \Sup &     A Unified Framework for Deep Symbolic Regression &        \cite{uDSR_DSRbased}  \\
\texttt{AIFeynman 2.0}  &            \Simp, \DA &      Symbolic regression exploiting graph modularity &           \cite{AIFeynman2}  \\
      \texttt{AFP\_FE}  &                   \GP &  AFP with co-evolved fitness estimates, Eureqa-esque &      \cite{EureqaPaper2009}  \\
          \texttt{DSR}  &                   \RL &                             Deep Symbolic Regression &          \cite{PetersenDSR}  \\
          \texttt{AFP}  &                   \GP &                      Age-fitness Pareto Optimization &  \cite{EureqaPaper2011_AFP}  \\
      \texttt{gplearn}  &                   \GP &             Koza-style symbolic regression in Python &              \cite{gplearn}  \\
     \texttt{GP-GOMEA}  &                   \GP &             GP-Optimal Mixing Evolutionary Algorithm &             \cite{GP_GOMEA}  \\
         \texttt{ITEA}  &                   \GP &                        Interaction-Transformation EA &                 \cite{ITEA}  \\
        \texttt{EPLEX}  &                   \GP &                        $\epsilon$-lexicase selection &                \cite{EPLEX}  \\
     \texttt{NeSymReS}  &                  \Sup &               Neural Symbolic Regression that Scales &  \cite{NeSymReS_EndToEndSR}  \\
       \texttt{Operon}  &                   \GP &                     SR with Non-linear least squares &               \cite{OPERON}  \\
        \texttt{SINDy}  &             \NeuroSym &         Sparse identification of non-linear dynamics &    \cite{Sindy_SymbolsInNN}  \\
       \texttt{SBP-GP}  &                   \GP &        Semantic Back-propagation Genetic Programming &               \cite{SBP_GP}  \\
          \texttt{BSR}  &                 \MCMC &                         Bayesian Symbolic Regression &      \cite{BSR_Bayesian_SR}  \\
         \texttt{FEAT}  &                   \GP &                  Feature Engineering Automation Tool &                 \cite{FEAT}  \\
          \texttt{FFX}  &                 \Rand &                             Fast function extraction &               \cite{FFX_SR}  \\
         \texttt{MRGP}  &                   \GP &              Multiple Regression Genetic Programming &                 \cite{MRGP}  \\
\bottomrule
\end{tabular}
\end{center}
\caption{Summary of baseline symbolic regression methods along with the the underlying techniques they rely on: reinforcement learning (\RL), genetic programming (\GP), problem simplification schemes (\Simp), end-to-end supervised learning (\Sup), dimensional analysis (\DA), neuro-symbolic / auto-differentiation based sparse fitting techniques (\NeuroSym), Markov chain Monte Carlo (\MCMC) and random search (\Rand).}
\label{table:baselines_summary}
\end{table*}

\definecolor{NoiselessViolet}{HTML}{7a2e70}
\def\properDSR{{\textcolor{NoiselessViolet}{\blacklozenge}}}

\begin{figure*}
\begin{center}
\includegraphics[angle=0, clip, width=\hsize]{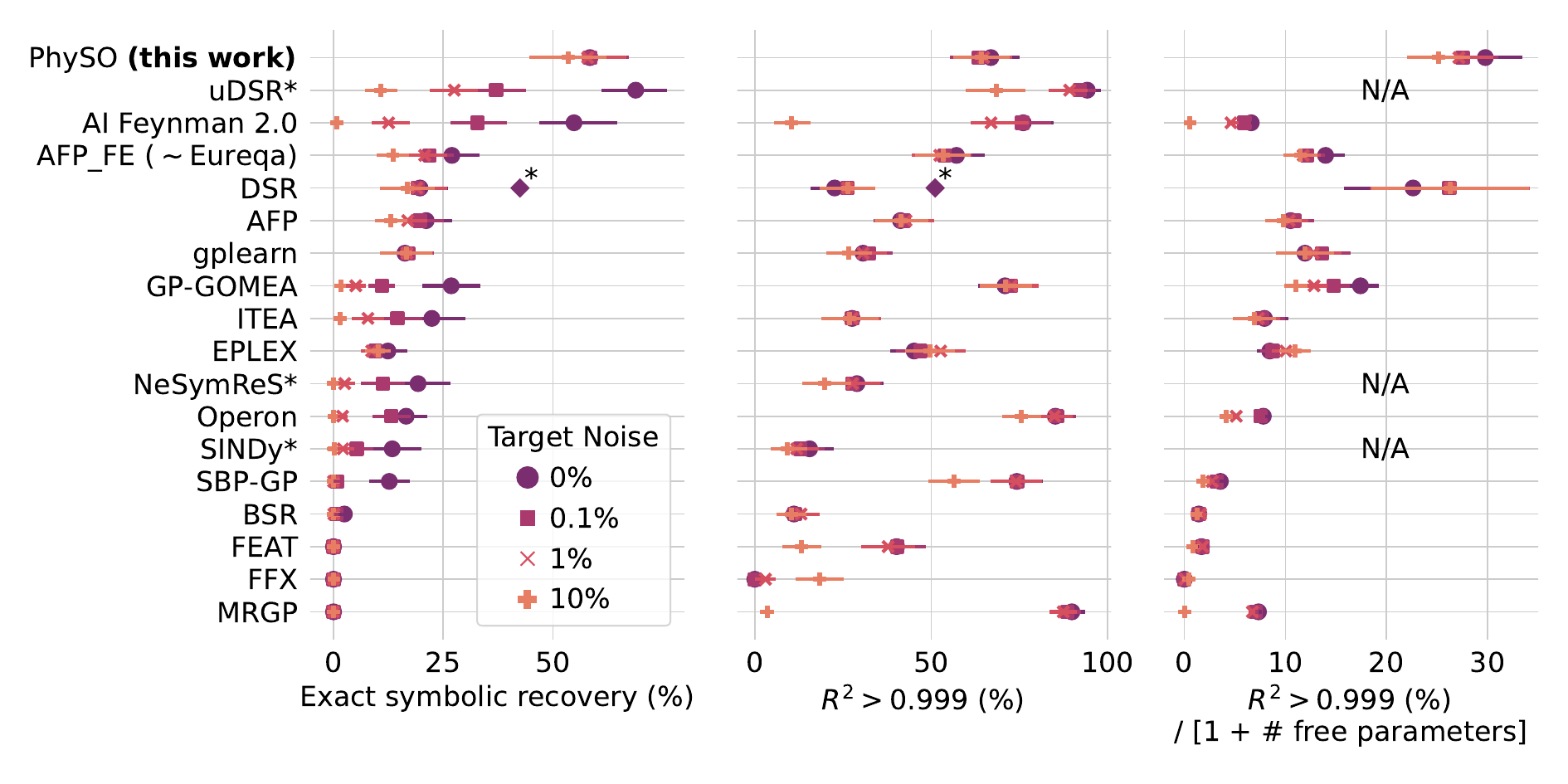}
\caption{Exact symbolic recovery rates, rates of accurate expressions (having $R^2>0.999$) and same rates normalized by the number of free parameters appearing in expressions for \texttt{PhySO} and other baseline methods on the Feynman benchmark. \texttt{PhySO} vastly outperforms all other methods in symbolic recovery in the presence of even minimal levels of noise ($> 0.1 \%$). In addition, the effectiveness of the dimensional analysis schemes of our \PhySO\ approach are clearly visible when comparing \texttt{DSR} (a purely RL method) with our implementation: \texttt{PhySO} (combining RL with dimensional analysis). Error-bars indicate a $95\%$ confidence interval, $\properDSR$ denotes performances of \texttt{DSR} \citep{SR_PG_improvements} reported in \cite{uDSR_DSRbased} on noiseless data with free constants allowed and * denotes that benchmarking conditions may vary and scores are polluted by approximately 5\% of results from another benchmark.}
\label{fig:feynman_benchmark}
\end{center}
\end{figure*}

\begin{figure}
\includegraphics[angle=0, clip, width=\hsize]{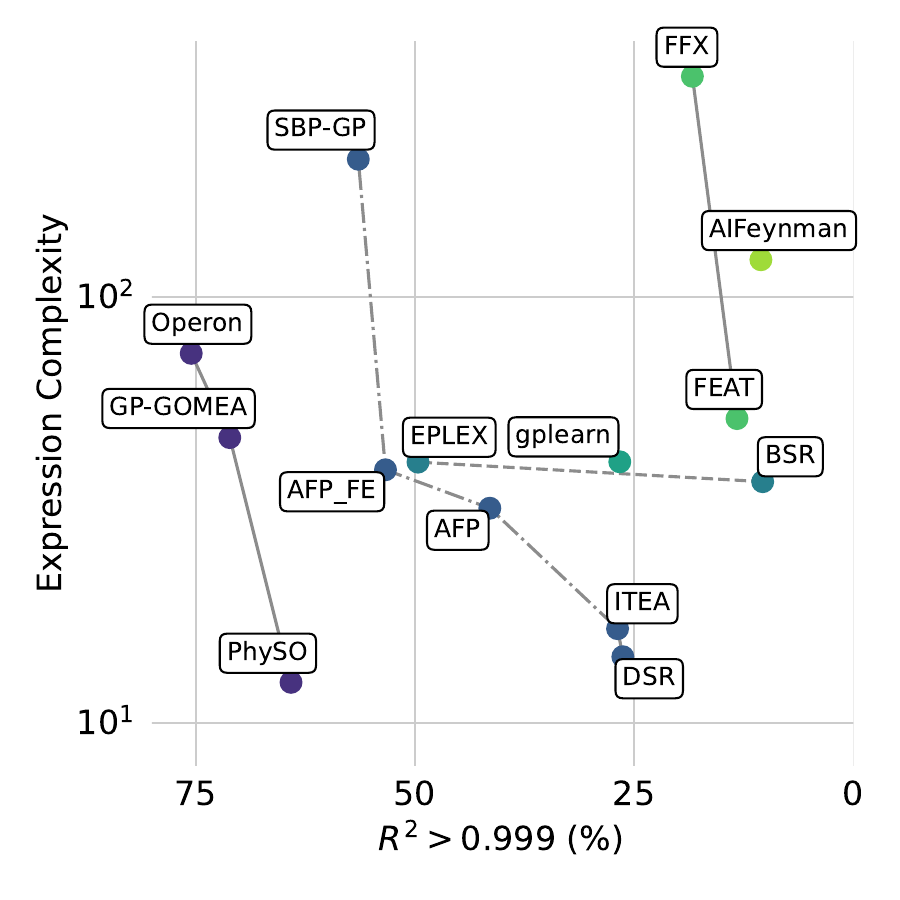}
\caption{Complexity versus rate of expressions having $R^2>0.999$ at a $10\%$ noise level for \texttt{PhySO} and other symbolic regression methods from the literature on the Feynman benchmark \citep{SRBench}. Lines and colors denote the 1\textsuperscript{st}, 2\textsuperscript{nd}, 3\textsuperscript{rd}, 4\textsuperscript{th}, 5\textsuperscript{th} and 6\textsuperscript{th} Pareto fronts following the \texttt{SRBench} algorithm comparison framework. \texttt{PhySO} is a Pareto optimum producing simple yet effective expressions.}
\label{fig:feynman_benchmark_pareto}
\end{figure}

\begin{figure*}
\begin{center}
\includegraphics[angle=0, clip, width=\hsize]{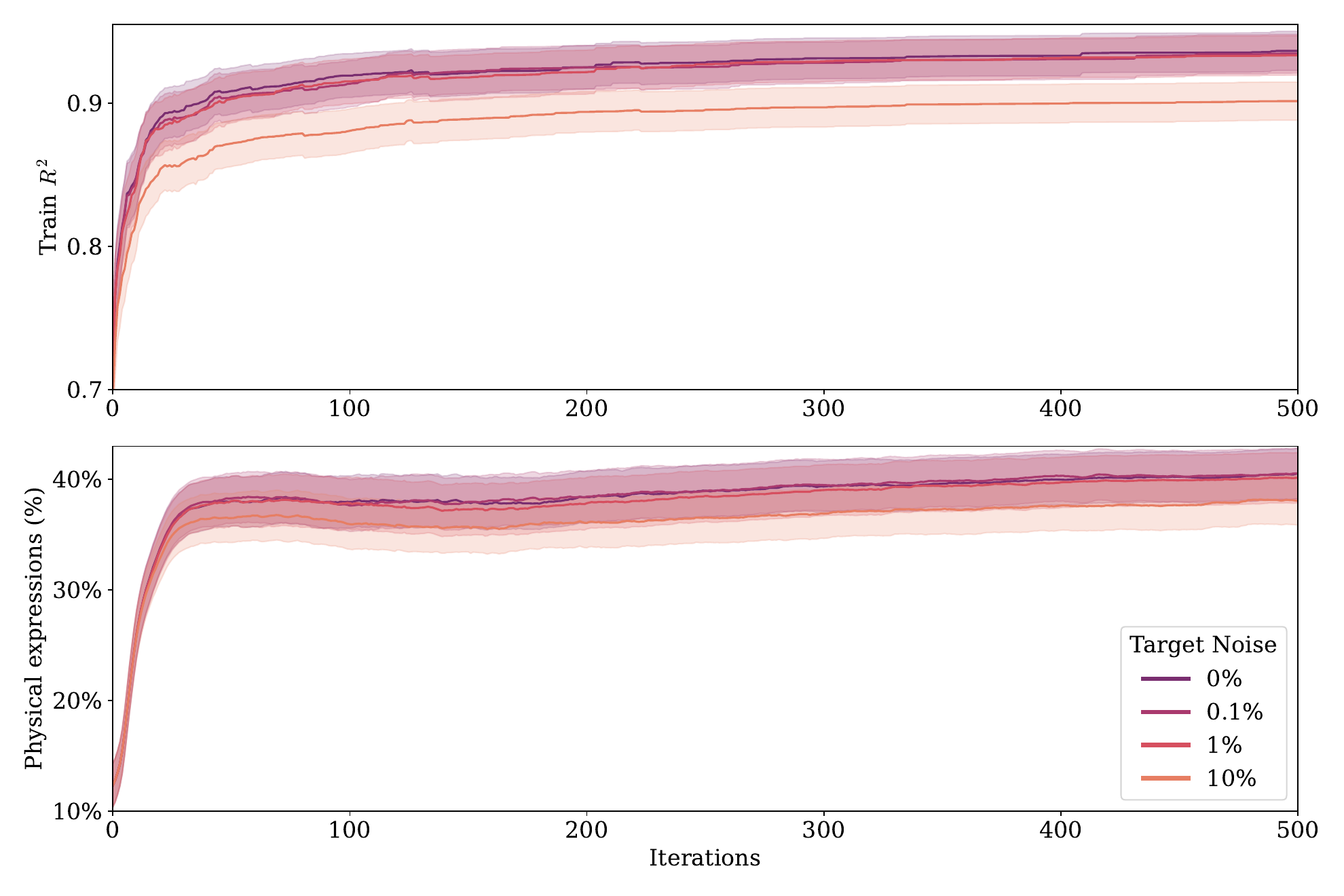}
\end{center}
\caption{$R^2$ value on training data and percentage of expressions natively proposed by the neural network that have balanced physical units averaged across the Feynman benchmark with error regions indicating a $95\%$ confidence interval. \PhySO's neural network learns to produce not only good fitting expressions but also physically meaningful ones.}
\label{fig:train_curve}
\end{figure*}

To validate the efficacy of our \PhySO\ method, we conducted benchmark tests using the widely-recognized Feynman symbolic regression benchmark. This set of challenges, first introduced by \cite{AIFeynman} and subsequently formalized in \texttt{SRBench} \citep{SRBench}, encompasses 120 equations including 100 sourced from the renowned Feynman Lectures on Physics \cite{FeynmanLectures} with the other 20 sourced from other textbooks: \citealt{GoldsteinClassicalMechanics, JacksonClassicalElectrodynamics, WeinbergGravitation, SchwartzQuantumFieldTheory}. The primary objective is to retrieve these equations using only the provided data points at various levels of noise.

Although this benchmark has inherent limitations, such as treating constants of nature (e.g., $G$, $c$, $\hbar$) and discrete physical values from quantum mechanics as continuously varying input variables (which places a higher emphasis on the implementation of the problem simplification schemes developed in \cite{AIFeynman}), it offers a comprehensive representation of the diversity of physical functional forms and remains a valuable standard for comparison as most SR methods have been thoroughly benchmarked on it (see \citealt{SRBench}).\\

Details on the benchmarking procedure can be found in \ref{subsec:feynman_protocol}. Results on exact symbolic recovery are provided in \ref{subsec:feynman_results_symbolic}, while findings regarding fit quality are presented in \ref{subsec:feynman_results_fit}. Finally, we provide training curves in \ref{subsec:learning_curves}.

\subsection{Benchmarking procedure}
\label{subsec:feynman_protocol}

We meticulously adhered to the established protocol delineated in \texttt{SRBench} by \cite{SRBench}, setting our \texttt{PhySO} algorithm to identify expressions that fit 10,000 data points corresponding to each Feynman benchmark equation. \texttt{PhySO} was only allowed to evaluate a maximum of one million expressions during each run and exact symbolic recovery was assessed by ensuring the difference between the expression generated by \texttt{PhySO} and the target expression reduced to a constant or that the fraction simplified to a constant using the \texttt{SymPy} library for symbolic mathematics \citep{sympy}. In addition, fit quality was assessed using the $R^2$ metric defined as $R^2 = 1 - \frac{\sum_{i=1}^N (y_i - f(\mathbf{x}_i))^2}{\sum_{i=1}^N (y_i - \bar{y})^2}$ on 100,000 noiseless test data points. As per benchmark rules, in order to ensure robustness, for each equation, this procedure was repeated multiple times (opting here for 5 trials over 10 due to the considerable computational demands associated with such benchmarks), each with a unique random seed, and the recovery rates were subsequently averaged. In alignment with \texttt{SRBench} stipulations, equations I.26.2, I.30.5, and test\_10 (containing $\arccos$ and $\arcsin$ functions) as well as II.11.17 were excluded from our results. The whole benchmark tests were conducted across four noise levels: $0\%$, $0.1\%$, $1\%$ and $10\%$, leading to the evaluation of 2,320,000,000 expressions.\\

We ran \texttt{PhySO} using the hyper-parameters and reward metric given in Section \ref{sec:method} (with the notable exception of the trigonometric prior which was set to a maximum nesting of one) and allowing the use of $\{ +, -, \times, /, 1/\placeholder, \sqrt{\placeholder}, \placeholder^2, -{\placeholder}, \exp, \log, \cos, \sin\}$\ as well as two dimensionless adjustable free constants and a constant equal to one $\{\theta_1, \theta_2, 1\}$. After each run, the first few expressions (in accuracy) of the Pareto front were inspected, which proved beneficial for cases where \texttt{SymPy} faced simplification challenges only and making a marginal difference of approximately 1\% in recovery rate. Notably, while the Feynman dataset includes unit information for each variable, \texttt{PhySO} is the only method that capitalizes on this feature since its introduction in \cite{AIFeynman}, a testament to its unique physics specific design. For the sake of reproducibility, we provide all the code required to execute the benchmark using \texttt{PhySO} as well as the detailed \texttt{SRBench}-style results regarding each run.\\

We compare the performance of our \PhySO\ approach to other SR algorithms with documented exact symbolic recovery rates, as reported in \citep{SRBench} and \citep{uDSR_DSRbased}. These algorithms are summarized in Table \ref{table:baselines_summary}. Remarkably, this includes \texttt{AFP\_FE} a \texttt{Eureqa}-like method, by the same authors combining \texttt{AFP} with \texttt{Eureqa}’s method for fitness estimation \citep{SRBench} and which we denote as \texttt{AFP\_FE} ($\sim$ \texttt{Eureqa}). In \citealt{SRBench}, \texttt{DSR} \citep{PetersenDSR} was not permitted to use any free parameters when generating expressions, greatly hindering its capabilities; we therefore also consider the performance of the latest version of \texttt{DSR} \citep{SR_PG_improvements} self-reported in the ablation study of \cite{uDSR_DSRbased} which relies on more suitable hyper-parameters as a baseline. However, we note that is important to exercise caution when interpreting this additional \texttt{DSR} performance data-point as well as the performances of \texttt{SINDy}, \texttt{NeSymReS}, and \texttt{uDSR} as our available data only offers their final scores on a composite dataset, which encompasses both the Feynman benchmark and the Strogatz benchmark \citep{Strogatz} - the latter accounting for approximately 5\% of the total score. This aggregated score is what we illustrate in our figures throughout this Section. In addition, it is worth noting that the exact conditions under which \texttt{SINDy} and \texttt{NeSymReS} were benchmarked are unknown and that in the case of \texttt{uDSR} and the additional \texttt{DSR} data-point, the benchmarking respectively permitted an evaluation of up to 2 million and 0.5 million expressions respectively, in contrast to the 1 million limit set for other methods. Furthermore, detailed results for these methods, in particular those regarding the specific expressions they identified, are unavailable, preventing their inclusion in our comparative analysis when concerning expression metrics (complexity or number of free parameters). Although, per \texttt{SRBench} rules, we permitted our method to evaluate up to 1 million expressions compared to \texttt{DSR}'s 0.5 million, \texttt{PhySO} typically identifies the correct expression well before reaching this limit or not at all. Additionally, while \texttt{DSR}'s $42\%$ score is influenced by another benchmark, the impact is very low, accounting for only $5\%$. This external benchmark is relatively straightforward, with \texttt{DSR} achieving around $25\%$ even without free parameters \citep{SRBench}, indicating its limited effect on the overall score. Thus, we believe a direct comparison between \texttt{PhySO}'s score and \texttt{DSR}'s from \citet{uDSR_DSRbased} is valid especially considering the gap in performance as detailed in the next sub-section.

\subsection{Exact symbolic recovery}
\label{subsec:feynman_results_symbolic}

Figure \ref{fig:feynman_benchmark} presents the performance of \texttt{PhySO} against baseline algorithms from Table \ref{table:baselines_summary} on the Feynman benchmark. This includes the average exact symbolic recovery rate, accurate expression rate (defined as those with a fit coefficient $R^2 > 0.999$), and normalized accurate expression rate considering the number of free parameters in the expressions, across different noise levels.
Compared to \texttt{DSR}, which strictly relies on reinforcement learning, \texttt{PhySO} utilizes both reinforcement learning and dimensional analysis. With \texttt{DSR}'s score at roughly $42\%$, our method's $58.5\%$ score highlights the significant benefits of incorporating dimensional analysis.
In the realm of physics, the exact symbolic recovery rate is a paramount metric and given that real-world physics data is often noisy, the resilience of an algorithm to noise is also crucial. However, with a minor noise level of $0.1\%$, many high-performing methods see their recovery rates almost halved. In contrast, \texttt{PhySO} maintains consistent performances. Remarkably, at a $10\%$ noise level, where most methods' recovery rates dip below $20\%$, and even high performers like \texttt{uDSR} and \texttt{AI Feynman 2.0} score only $10.7\%$ and $0.7\%$ respectively, \texttt{PhySO} continues to accurately recover expressions over $53\%$ of the time.\\

In noiseless scenarios \texttt{PhySO} is only surpassed by \texttt{uDSR} which relies on a cocktail of five of the most potent SR techniques: reinforcement learning for iterative adjustments, genetic programming for enhanced randomization and exploration, supervised learning to leverage existing knowledge, neuro-symbolic style sparse coefficient fitting for its linear symbolic modules and powerful simplification strategies, similar to those utilized by \texttt{AI Feynman 2.0}, which narrowly lags behind \texttt{PhySO}. These techniques rely on the exploitation of separability (\eg simplifying the search of $f(x_1, x_2)$ to the search of the simpler functions $f_1(x_1)$ and $f_2(x_2)$ with $f(x_1, x_2) = f_1(x_1) + f_2(x_2)$), symmetry (\eg simplifying the search of $f(x_1, x_2)$ to the search of $f_1(x_1, x_2)$ and $f_2(x_2)$ with $f(x_1, x_2) = f_1(x_1, f_2(x_2))$), and many other schemes to circumvent the intricate functional forms in the benchmark. Despite relying solely on reinforcement learning and dimensional analysis, on noiseless data \texttt{PhySO} rivals \texttt{uDSR} and surpasses \texttt{AI Feynman 2.0}, demonstrating the effectiveness of our approach.\\

It is worth noting that while the aforementioned \texttt{AI Feynman}-style ``divide and conquer'' simplification strategies are effective, they are extremely noise sensitive, a scenario where \texttt{PhySO}'s approach remains stable. In summary, this benchmark shows that incorporating dimensional analysis constraints into SR significantly bolsters performance. Given the improvements shown from \texttt{PhySO} over \texttt{DSR} thanks to the inclusion of units constraints, and given \texttt{uDSR}'s impressive performances in noiseless scenarios, we believe combining \PhySO\ with \texttt{uDSR} could elevate outcomes even further.

\subsection{Fit quality}
\label{subsec:feynman_results_fit}

Regarding the fraction of expressions with an $R^2>0.999$, many methods achieve high scores by incorporating an extensive number of free constants, resulting in intricate expressions that often lack interpretability and are nonsensical from a dimensional analysis standpoint. For example, \texttt{AI Feynman 2.0}, when not identifying the precise symbolic expressions, tends to generate complex expressions comprising, post-simplification, an average of 147 symbols and 18 free constants due to its brute-force polynomial fitting approach. Similarly \texttt{Operon} \footnote{It should be noted that a recent improvement of Operon (see \cite{OPERON_simpler}) allowing it to produce simpler expressions was introduced after the first submission of our paper. We expect this improved version to perform better on the Feynman benchmark.} and \texttt{MRGP} expressions contain on average respectively 17 and 88 free constants post-simplification at a $10\%$ noise level. This is not a problem in many fields where human-interpretability is not a priority. However, given the importance of this criterion in physics we also show in Figure \ref{fig:feynman_benchmark} the rate of accurate expressions normalized by the number of free constants plus 1. \texttt{PhySO} emerges as the leading method in generating succinct, physically coherent, and interpretable expressions that best approximate a dataset, that is when it is not able to recover the exact underlying expression all together.

This is further illustrated in Figure \ref{fig:feynman_benchmark_pareto}, where we show Pareto frontiers of expression complexity versus fit quality at a $10\%$ noise level for all benchmarked methods with available output expression information. On this plot \texttt{PhySO} is a Pareto optimum demonstrating its ability to produce simple yet good-fitting expressions.

\subsection{Learning curves}
\label{subsec:learning_curves}

Due to its very constraining nature, using a yet untrained neural network, our \insitu units prior often conflicts with the length prior which is essential to avoid the expression generation phase going on forever. This typically results in the majority of expressions being discarded due to this conflict during the first iterations of the training process. However, enabling the neural network to learn on physically correct expressions, and enabling it able to observe local units constraints, allows it to actively learn dimensional analysis rules. This is shown in Figure \ref{fig:train_curve}, which gives the fraction of physical expressions successfully generated over iterations of learning averaged over all runs of the Feynman benchmark at each level of noise.

Moreover, Figure \ref{fig:train_curve} presents the evolution of the $R^2$ fit coefficient on training data for the best expression identified at each iterations. The figure demonstrates that as the iterations progress, the neural network not only improves in generating expressions with better fits but also refines its capacity to produce expressions that are physically meaningful.

In our observations, while \PhySO\ occasionally escapes local minima through stochastic variations, convergence is typically characterized by the neural network mostly producing identical expressions. This state of convergence is typically reflected by both average fit quality and rate of physical expression remaining static, as well as by the reward distribution peaking. The rate of convergence is dependent on the difficulty of the case, the level of noise and the chosen hyper-parameters. As depicted in Figure \ref{fig:train_curve}, under the hyper-parameters detailed in this study, \PhySO\ typically reaches convergence well within several hundred iterations. Note that since it is operating in a reinforcement learning framework, \PhySO\ is trained on `moving targets' as its targets consist of expressions generated by itself during the last iteration which is characterized by the loss not decreasing during training except when it starts consistently producing similar equations while converging.


\section{Astrophysical case studies}
\label{sec:case_studies}

\def\CompleteLibrary{{$\{ +, -, \times, /, 1/\placeholder, \sqrt{\placeholder}, \placeholder^2, \exp, \log, \cos, \sin, 1 \}$\ }}

\def\MaxTrialExpressions{10 million}
\def\mColored{\textcolor{myred}{m}}
\def\vColored{\textcolor{myred}{v}}
\def\cColored{\textcolor{myddblue}{c}}
\def\RelatEnergy{{$E = \frac{\mColored \cColored^2}{\sqrt{1 -\vColored^2/\cColored^2}}$\ }}
\def\EColored{\textcolor{myred}{E}}
\def\LColored{\textcolor{myred}{L}}
\def\GMColored{\textcolor{myddblue}{G M}}
\def\bColored{\textcolor{myddblue}{b}}
\def\IsochroneAction{{$J_r = \frac{\GMColored}{\sqrt{-2 \EColored}} - \frac{1}{2} \Big( \LColored + \frac{1}{2} \sqrt{\LColored^2 - 4 \GMColored \bColored} \, \Big)$\ }}
\def\rColored{\textcolor{myred}{r}}
\def\rhoColored{\textcolor{myddblue}{\rho_0}}
\def\RsColored{\textcolor{myddblue}{R_s}}
\def\NFWprofile{{$\rho = \rhoColored / \Big( \frac{\rColored}{\RsColored} (1 + \frac{\rColored}{\RsColored})^2 \Big)$\ }}
\def\tColored{\textcolor{myred}{t}}
\def\alphahoColored{\textcolor{myddblue}{\alpha}}
\def\PhiColored{\textcolor{myddblue}{\Phi}}
\def\fColored{\textcolor{myddblue}{\omega}}
\def\DHO{{$y = e^{-\alphahoColored \tColored} \cos (\fColored \tColored + \PhiColored) $\ }}
\def\moColored{\textcolor{myred}{m_1}}
\def\mtColored{\textcolor{myred}{m_2}}
\def\rColored{\textcolor{myred}{r}}
\def\GColored{\textcolor{myddblue}{G}}
\def\ClassicalGravity{{$F = \frac{\GColored \moColored \mtColored}{\rColored^2}$\ }}
\def\xColored{\textcolor{myred}{x}}
\def\OmegaColored{\textcolor{myddblue}{\Omega_m}}
\def\HColored{\textcolor{myddblue}{H_0}}
\def\ExpansionLaw{{$H^2(x\equiv1+z) = \HColored^2 (\OmegaColored \xColored^3 + (1 - \OmegaColored))$\ }}

We now showcase our \PhySO\ method on a panel of astrophysical test cases: the relativistic energy of a particle is examined in sub-section \ref{subsec:case_relativity}, the law describing the expansion of the Universe in sub-section \ref{subsec:case_cosmo}, the isochrone action from galactic dynamics in sub-section \ref{subsec:case_galactic} and additional toy test cases given in \ref{subsec:case_additional}. We give the results along with an ablation study, disabling specific components our system to determine their impact on performance, in \ref{subsec:astro_results}. We perform this ablation study in a noiseless scenario using mock data but still demonstrate \PhySO's abilities on observational noisy data for the case detailed in sub-section \ref{subsec:case_cosmo}, showing that the method can successfully recover physical laws and relations from real or synthetic data. Mock data generation details are given in Appendix \ref{appendix:astro_data} along with units of all variables and constants involved. Note that for each of these showcases, we explicitly add the free constants described in Appendix \ref{appendix:astro_data} along with their units in \PhySO's library of available tokens. We use the hyper-parameters and and reward metric detailed in Section \ref{sec:method} and limit ourselves to the exploration of \MaxTrialExpressions\ trial expressions which roughly takes $\sim 4$ hours (using all cores of the systems shown in Figure \ref{fig:physo_parallel_efficiency}) and is only necessary for the most difficult case (the relativistic energy). 
In addition, for the relativistic energy showcase, we give a Pareto front which shows the most accurate expression based on RMSE (root mean squared error) for each level of complexity.
Moreover, similarly to the benchmarking in Section \ref{sec:feynman}, we define the successful exact symbolic recovery of an expression by its symbolic equivalence using the \texttt{SymPy} symbolic simplification subroutine \citep{sympy}. Finally, we agnostically rely on the same library of choosable tokens for all test cases: \CompleteLibrary to which we only add input variables and free or fixed constants depending on the test cases.

\begin{figure*}
\begin{center}
\includegraphics[angle=0, clip, width=\hsize]{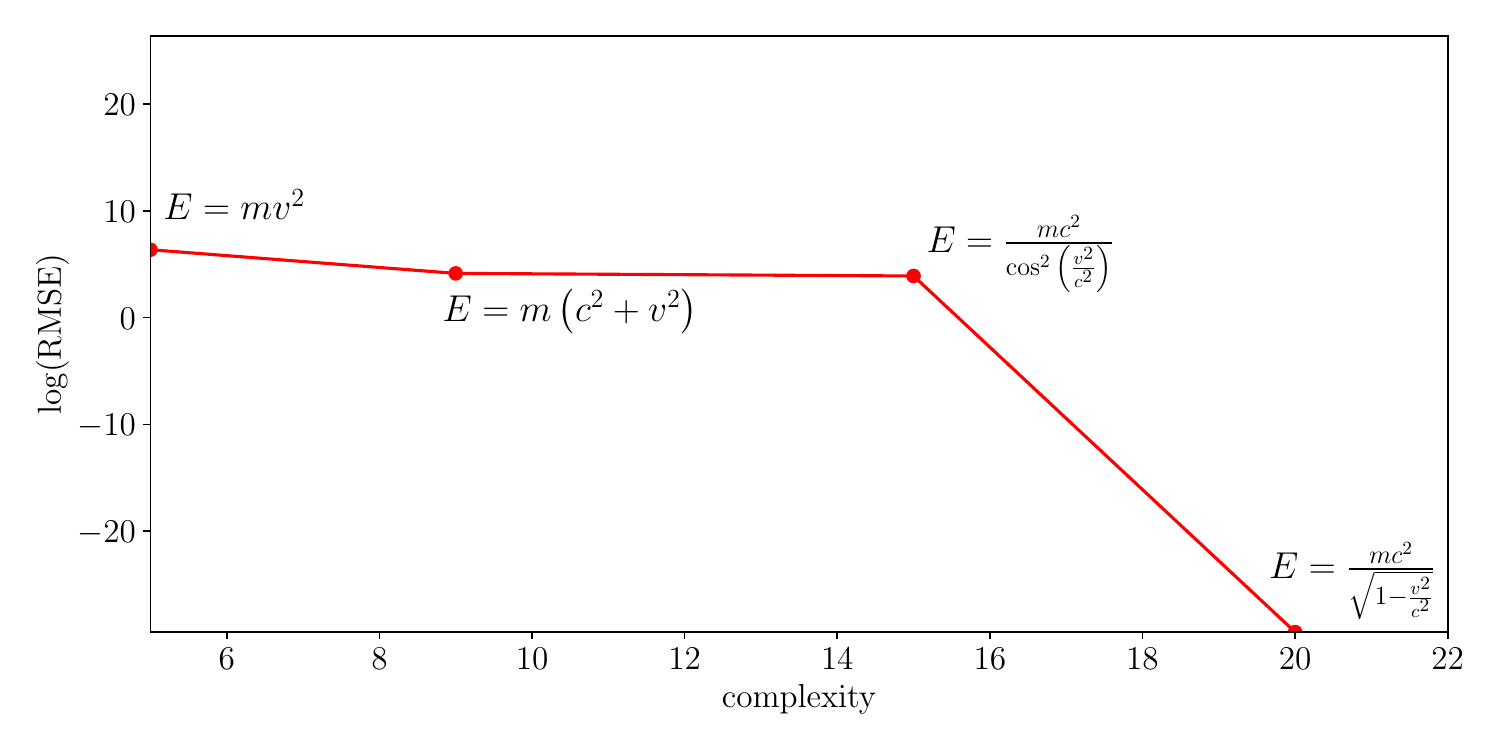}
\end{center}
\caption{Pareto-front encoding accuracy-complexity trade off of recovered physical formulae typically recovered using our \PhySO\ method when applied to data for the relativistic energy of a particle. We recover the relativistic expression as well as the classical approximation. Note that although the exact classical expression $\frac{1}{2}mv^2$ is encountered by \PhySO\ it is Pareto dominated by the simpler $mv^2$ expression.}
\label{fig:relatenergy_pareto}
\end{figure*}

\subsection{Relativistic energy of a particle}
\label{subsec:case_relativity}

Let us consider the expression for the relativistic energy of a particle:
\begin{equation}
    E = \frac{mc^2}{\sqrt{1-\frac{v^2}{c^2}}} \, ,
\end{equation}
where $m$, $v$ and $c$ are respectively the mass of the particle, its velocity and the speed of light.

Using the aforementioned library of tokens as well as the $\{m, v\}$ input variables and a free constant $\{c \}$, \PhySO\ is able to successfully recover this expression $100 \%$ of the time. Figure \ref{fig:relatenergy_pareto} contains the Pareto front of recovered expressions where similarly to \cite{AIFeynman2}, we showcase that we are able to recover the relativistic energy of a particle as well as the classical approximation which has a lower complexity.

However, we note that our system is able to recover the exact expression for the relativistic energy test case without any of the powerful simplification on which relies the \texttt{AI Feynman 2.0} approach proposed in \citep{AIFeynman2} (in particular, the identification of symmetries as well as the identification of additive and multiplicative separability), nor by simplifying the problem further by treating $c$ (a constant of nature) as a variable taking a range of different values as in \citep{AIFeynman2}. Neither \texttt{DSR} \citep{SR_PG_improvements} nor \texttt{AI Feynman} \citep{AIFeynman2} are able to crack this case under these more stringent conditions.

\subsection{Expansion of the Universe}
\label{subsec:case_cosmo}

The next case study we examine is the Hubble Diagram of supernovae type Ia, namely the change in the observed luminosity of these important standard candles as a function of redshift $z$. This is one of the major pieces of evidence that indicates that the Universe is experiencing an accelerating expansion, and it is also one of the observational pillars underlying $\Lambda$ Cold Dark Matter ($\Lambda$CDM) cosmology in which Dark Energy dominates the energy-density budget of the Universe. 

\begin{figure}
\begin{center}
\includegraphics[angle=0, clip, width=\hsize]{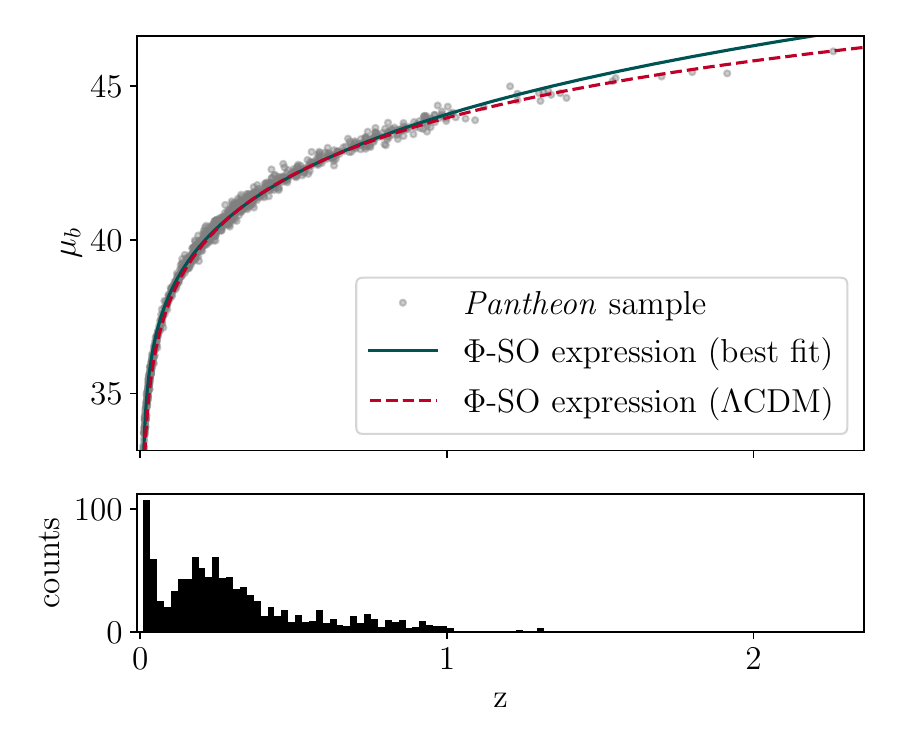}
\end{center}
\caption{Symbolic regression results when applying the \PhySO\ algorithm (allowing two free parameters) to the Hubble Diagram of supernovae Ia from the {\it Pantheon} sample. \PhySO\ rediscovers the $\Lambda$CDM relation (in red) as well as another relation (in blue) which has a slightly better fit than $\Lambda$CDM when solely considering {\it Pantheon}'s observational constraints due to the over-abundance of low z SNe.}
\label{fig:Hubble_Diagram}
\end{figure}

We will use the so-called {\it Pantheon} state-of-the-art compilation dataset \citep{Scolnic2018}, shown in Figure~\ref{fig:Hubble_Diagram}. We use a similar calibration and follow an almost identical methodology as \citet{Exhaustive_SR}, to find the Hubble parameter $H(z)$ from the measured supernova magnitude and redshift pairs. Following \citet{Exhaustive_SR}, we use the auxiliary function 
\begin{equation}
y(x \equiv 1+z) \equiv H(z)^2 \, ,
\end{equation}
which for $\Lambda$CDM in a flat Universe with negligible radiation pressure is
\begin{equation}
y_{\Lambda {\rm CDM}}(x) = H_0^2 (\Omega_m x^3 + (1 - \Omega_m)) \, ,
\label{eq:hubble_target}
\end{equation}
where $\Omega_m$ is the matter density parameter and $H_0$ is the Hubble constant. In a flat Universe model the cosmological luminosity distance is
\begin{equation}
d_L(z)=(1+z) \int_0^z \frac{c \, dz^\prime}{H(z^\prime)} \, ,
\label{eqn:luminosity-distance}
\end{equation}
where $c$ is the speed of light.

\def\Hparam{\widetilde{H}}
\begin{table*}[]
\begin{center}
\begin{tabular}{lcccc}
\hline
Expression                                                     & Complexity & $\Hparam$  & $\theta$ & $R^2$  \\ \hline
$\Hparam^{2} \sqrt{\theta^{2} + \log{\left(\theta + x \right)}}$ & 14         & 5.175  & -0.01    & 0.9955 \\
$\Hparam^{2} \sqrt{\theta + x}$                                  & 9          & 4.692  & -1.01    & 0.9946 \\
$\Hparam^{2} \log{\left(x \right)}$                              & 6          & 7.499  & -        & 0.9627 \\
$\Hparam^{2} \log{\left(x \right)}^{2}$                          & 8          & 28.276 & -        & 0.9523 \\
$\Hparam^{2} \left(\theta x^{3} - \theta + 1\right)$   \footnote{$\Lambda$CDM expression for reference}              & 14         & 73.3   & 0.315    & 0.9166 \\ \hline
\end{tabular}
\end{center}
\caption{Pareto accuracy-complexity trade-off expressions (for the auxiliary function $y(x \equiv 1+z) \equiv H(z)^2$) applying the \PhySO\ algorithm (allowing two parameters) to the Hubble Diagram of supernovae Ia from the {\it Pantheon} sample. Although \PhySO\ generates the $\Lambda$CDM expression, it is not a Pareto optimum when solely considering {\it Pantheon}'s observational constraints due to the over-abundance of low z SNe. We include it for reference as the last line of this Table.}
\label{table:pareto_expansion}
\end{table*}

We adapt our machinery to the Hubble diagram problem by integrating numerically the $H(z \equiv x-1)=\sqrt{y(x)}$ functions proposed by the algorithm under Eqn.~\ref{eqn:luminosity-distance} the implied luminosity distance $d_L$. These are then trivially converted into a distance modulus $\mu(z) = 5 \log_{10} ( d_L(z)/10 \pc )$, which we compare to the {\it Pantheon} data following the procedure given in sub-section \ref{subsec:learning}.

This Hubble Diagram example showcases the capability of the software to include free ``constants'' (here we include one having the units of $H_0$ and the other being dimensionless as $\Omega_m$) in the expression search, whose values are found thanks to auto-differentiation via L-BFGS optimization, as mentioned in Section~\ref{subsec:learning}. The optimal values of these constants need to be calculated after being passed through the numerical integration step (integrating Eqn.~\ref{eqn:luminosity-distance} via \texttt{PyTorch} differentiable cumulative trapezoids), which turns out to be the main bottleneck of the problem in terms of computational cost. However, this also shows that the algorithm allows one to derive expressions that are subsequently passed through complicated operations before being compared to data.

The Pareto front is given in Table \ref{table:pareto_expansion} alongside the $\Lambda$CDM expression. Although we are able to recover it using synthetic data, we note that as \cite{Exhaustive_SR}, using observational data our system finds more accurate solutions at lower complexities than the $\Lambda$CDM model. 
Although this could signify that the $\Lambda$CDM theory is inaccurate, here we refrain from jumping to this conclusion because our system is only given the chance to confront its trial model of $H(z)$ to a relatively noisy dataset of standard candles where there is an over abundance of low z events, and is not provided other observational constraints such as the cosmic microwave background which might tilt the balance in favor of $\Lambda$CDM as the most accurate model at its level of complexity. However, although the $\Lambda$CDM expression is not the global minimum with this set of observational constraints, while exploring a space of increasingly accurate expressions our system recognizes it as an intermediate step, recording it in its history, before eventually converging to a different expression.
In addition, we note that it is not surprising that our system recovers the $\Lambda$CDM expression as we allowed a maximum of two free parameters since the main goal was simply to demonstrate our system's capabilities. We defer multi-parameter studies to future contributions.
Finally we are able to recover this expression by typically exploring $< 50$k expressions (which takes less than a minute on the systems shown in Figure \ref{fig:physo_parallel_efficiency}), the same order of magnitude as in the exhaustive symbolic regression approach proposed in \cite{Exhaustive_SR} but allowing more functions ($\cos, \sin, \exp, \log$).\\

\subsection{Isochrone action from galactic dynamics}
\label{subsec:case_galactic}

Another interesting application of symbolic regression is to derive perfect analytical properties of analytical models of physical systems. To this end, we chose to attempt to find the radial action $J_r$ of the spherical isochrone potential.
\begin{equation}
\Phi(r) = - \frac{G M}{b + \sqrt{b^2 + r^2}} \, ,
\end{equation}
where $G$ is the gravitational constant, $M$ is the mass of the model, $b$ is a length scale of the model, and $r$ is a spherical radius \citep{GalacticDynamics}. Action variables are special integrals of motion in integrable potentials which can be used to describe the orbit of an object in a system, and they are of particular interest in Galactic Archaeology as they are adiabatic invariants, so they are preserved if a galaxy or stellar system has evolved slowly. The isochrone is the only potential model to have actions known in analytic form in terms of elementary functions\footnote{We have recently shown that actions can be calculated numerically from samples of points along orbits in realistic galaxy potentials using deep learning techniques \citep{ActionFinder}.}. For the case of the isochrone model, the radial component of the action of a particle can be expressed as
\begin{equation}
J_r = \frac{G M}{\sqrt{-2E}} - \frac{1}{2} \Big( L + \frac{1}{2} \sqrt{L^2 - 4G M b} \Big) \, ,
\label{eqn:Jr}
\end{equation}
where $E$ and $L$ are, respectively, the particle energy and total angular momentum \citep{GalacticDynamics}.

We provide our algorithm numerical values of $J_r$ (which has units of angular momentum) given $L$ and $E$, and leave $b$ as a free scaling parameter. Since we expect each occurrence of $M$ to be accompanied by an occurrence of the gravitational constant, we provide the algorithm with $G M$ as a single variable.

This expression (Eqn.~\ref{eqn:Jr}) could not be solved either by the standard \texttt{DSR} algorithm \citep{SR_PG_improvements}, or by the \texttt{AI Feynman 2.0} algorithm \citep{AIFeynman}. Our algorithm was also not able to identify the equation in \MaxTrialExpressions\ guesses. However, one of the steps of the \texttt{AI Feynman 2.0} algorithm is a test for additive and multiplicative separability of the mystery function, and it creates new datasets for each separable part. For the case of additive separability, the units remain unchanged, and so it is trivial to simply provide our \PhySO\ algorithm separated data generated by \texttt{AI Feynman 2.0} to be fitted in turn, one at a time. Thus the first term of the right hand side of Eqn.~\ref{eqn:Jr} (with an $E$ dependence) was easily solved together with a fitted additive free constant. We then subtracted the fitted constant from the second dataset, and \PhySO\ correctly recovered the second term on the right hand side of Eqn.~\ref{eqn:Jr} (with an $L$ dependence).

\subsection{Supplementary cases}
\label{subsec:case_additional}

In addition to the cases above, we consider the following set of textbook equations for the ablation study in \ref{subsec:astro_results}. We include Newton's law of universal gravitation:
\begin{equation}
F = \frac{G m_1 m_2}{r^2} \, ,
\end{equation}
where $G$ is the universal gravitational constant, $m_1$ and $m_2$ are the masses of the attracting bodies and $r$ is the distance separating them. For this test case, we use $\{ m_1, m_2, r\}$ as input variables and leave $G$ as a free constant.

We also include a damped harmonic oscillator which appears in a wide range of (astro)-physical contexts:
\begin{equation}
y = e^{-\alpha t} \cos (\omega t + \Phi) \, ,
\label{eqn:damped_oscillator}
\end{equation}
where $\alpha$ and $\omega$ are respectively the damping parameter and the angular frequency of oscillations (both homogeneous to the inverse of a time) and $\Phi$ is the (dimensionless) phase. We leave these three parameters as free constants and use $t$ as our input variable.

Finally, we consider a Navarro–Frenk–White (NFW) halo profile \citep{NFWprofile} which is an empirical relation that describes the density profile $\rho(r)$ of halos of collisionless dark matter in cosmological N-body simulations:
\begin{equation}
\rho = \frac{\rho_0}{\frac{r}{R_s} \Big(1 + \frac{r}{R_s} \Big)^2} \, ,
\end{equation}
where $r$ is the radius which we use as an input variable and $\rho_0$ and $R_s$ are respectively the density and radius scale parameters which we leave as free constants.

\subsection{Ablation study}
\label{subsec:astro_results}

\begin{table*}[]
\begin{center}
\begin{tabular}{ll@{\hskip 1in}cccccc}
\multicolumn{2}{l}{Ablation configuration}                 & A \footnote{Full \PhySO\ method.} & B              & C              & D \footnote{Similar to \cite{SR_PG_improvements}.} & E              & F  \footnote{Solely relying on a random number generator.}             \\ \hline
\multicolumn{2}{l}{Physical units prior}                   & \checkmark                       &                & \checkmark     &                                                   & \checkmark     &                 \\
\multicolumn{2}{l}{Physical units informed neural network} & \checkmark                       & \checkmark     &                &                                                   &                &                 \\
\multicolumn{2}{l}{Neural network enabled}                 & \checkmark                       & \checkmark     & \checkmark     & \checkmark                                        &                &                 \\ \hline
                              &                            &                                  &                &                &                                                   &                &                 \\
Expression                    & \# expressions             &                                  &                &                &                                                   &                &                 \\ \hline
{\RelatEnergy}                & $10$M                      & 100 \%                           & 0 \%           & 60 \%          & 0 \%                                              & 20 \%          & 0  \%           \\
{\IsochroneAction}            & $4$M                       & 100 \%                           & 0 \%           & 80 \%          & 0 \%                                              & 60 \%          & 0  \%           \\
{\NFWprofile}                 & $2$M                       & 100 \%                           & 100 \%         & 40 \%          & 100 \%                                            & 20 \%          & 100  \%         \\
{\DHO}                        & $1$M                       & 100 \%                           & 0 \%           & 0 \%           & 0 \%                                              & 0 \%           & 0  \%           \\
{\ClassicalGravity}           & $100$K                     & 100 \%                           & 80 \%          & 100 \%         & 20 \%                                             & 80 \%          & 0  \%           \\
{\ExpansionLaw }              & $100$K                     & 100 \%                           & 100 \%         & 100 \%         & 100 \%                                            & 40 \%          & 40  \%          \\ \hline
                              & \textbf{Average}           & \textbf{100 \%}                  & \textbf{47 \%} & \textbf{63 \%} & \textbf{37 \%}                                    & \textbf{37 \%} & \textbf{23  \%}
\end{tabular}
\caption{Exact symbolic recovery rate summary and ablation study on our panel of astrophysical examples using noiseless synthetic data, averaged across 5 runs. By studying the performance in combinations of ablations of the \insitu units prior, the neural networks's ability to be informed of local units constraints, and of the neural network itself (\ie replaced by a random number generator when not marked as enabled), we show that all three are essential ingredients of the success of our \PhySO\ method. Input variables and free parameters are colored in red and blue respectively with fixed constants left in black.}
\label{table:results_summary_ablation}
\end{center}
\end{table*}

\begin{center}
\begin{figure}
\includegraphics[angle=0, clip, width=\hsize]{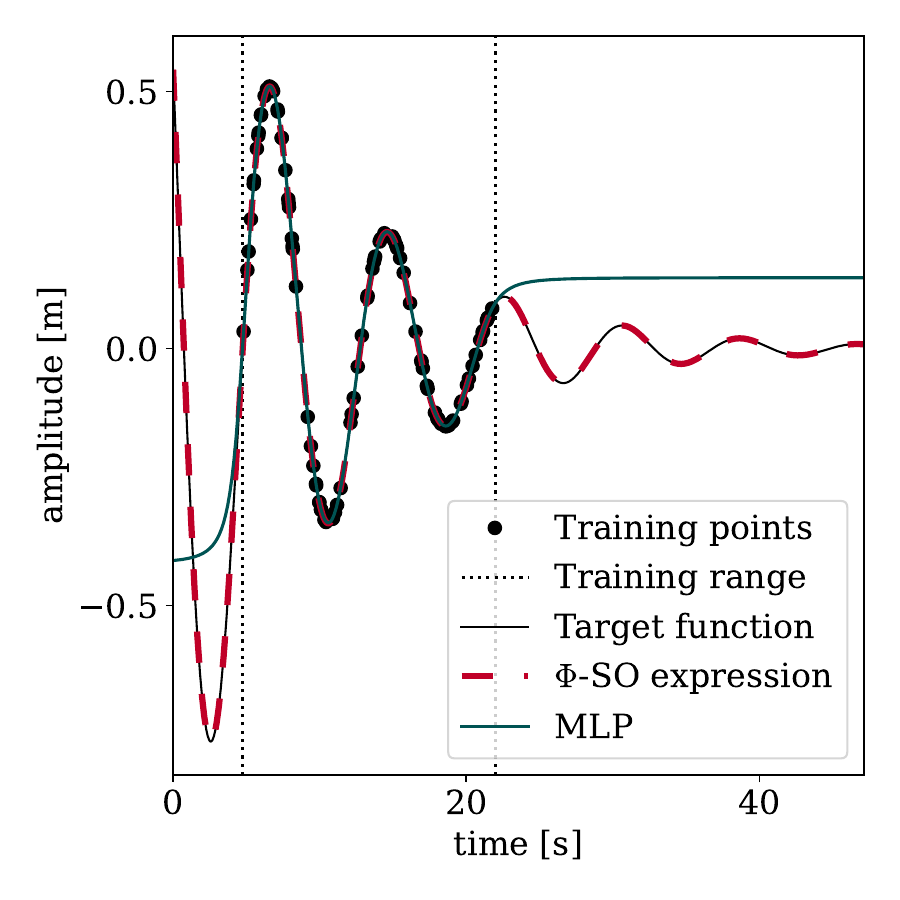}
\caption{Example of the generalization capability of SR. Here we show randomly drawn data points (black dots) from the damped harmonic oscillator model given in Eqn. \ref{eqn:damped_oscillator} (black line). The data are well fitted by an MLP (green line), which however fails in regions beyond the range of the training data (vertical dotted lines). In contrast, our SR algorithm \PhySO\ (red dashed line) manages to provide much more reliable extrapolation.}
\label{fig:generalization_demo}
\end{figure}
\end{center}

In physics, we often seek to build approximate models, such as might be obtained via a polynomial function or a Fourier series fit to some data. In those instances, the root mean square error is usually the criterion of relevance to determine whether the procedure worked well or not. However, here we wish to recover the ``true'' underlying model, in which case the recovery rate should be the criterion of success.

The performance of \PhySO\ on noiseless mock data from the test cases detailed above is summarized in the ablation study reported in Table \ref{table:results_summary_ablation}. There we also report SR performance after disabling the units prior (only using the units informed RNN), disabling the RNN's ability to be informed of local units units constraints (only using the units prior and a standard SR RNN), disabling both the units prior and units information (only using a standard RNN which is similar to the \citealt{SR_PG_improvements} setup), doing a units guided random search by using a random number generator {\it in lieu} of the RNN, and finally doing a purely random search. 

We show that merely constraining the choice of symbols using the external units prior distribution scheme (described in \ref{subsec:units_prior}) is not enough to ensure perfect symbolic recovery of physical laws, but that informing the RNN of local units constraints (as described in \ref{subsec:embedding}) is essential as it allows the RNN to actively learn units rules. In addition, we show that our system does not only rely on a mere brute force approach combined with units constraints, but that the deep reinforcement learning setup described in \ref{subsec:learning} is an essential ingredient of the success of \PhySO. 

It should be noted that in the NFW test case, simply expressing the inverse of a third-degree polynomial is sufficient to solve the problem. However, using the units prior without enabling the RNN to observe local units constraints or utilizing the units prior in conjunction with a random number generator can result in a lower recovery rate compared to the use of a standalone random number generator. This is due to the highly restrictive nature of the units prior which in a simple case like this can actually slow down the convergence toward the solution.

Finally, we also illustrate the generalization capabilities offered by virtue of finding the exact analytical expression 
underlying a dataset compared to a good approximation in Figure \ref{fig:generalization_demo}, where we show that such analytical expressions, as expected, vastly outperform a multilayer perceptron (MLP) neural network (here a 5 layers of 32 units MLP having sigmoid activations and being trained until convergence on a test set, following a mean squared error loss function at $10^-3$ learning rate using an Adam optimizer, \citealt{adam_optimizer}).


\section{Discussion}
\label{sec:discussion}

Since the Deep Symbolic Regression framework \citep{PetersenDSR} and most other SR methods work by maximizing fit quality, there are few constraints on the arrangement of symbols. However, the paths in fit quality and the paths in symbol arrangement toward the global minima (perfect fit quality and perfect symbol arrangement) are not necessarily correlated. This results in the curse of accuracy guided SR, as small changes in fit quality can hide dramatic changes in functional form and vice-versa. In essence, one can improve fit quality of candidates over learning iterations while getting further away from the correct solution in symbolic arrangement. Therefore strong constraints on the functional form, such as the one we are proposing in our setup, are of great value for guiding SR algorithms in the context of physics. This is an advantage that physics has and that \PhySO\ leverages by: (i) reducing the search space and (ii) enabling the neural network to actively learn dimensional analysis rules and leverage them to explore the space of solutions more efficiently. Although the possibility of making a physical units prior was hinted by \citet{DSR_priors}, to the best of our knowledge such a framework was never built before.\\



The guidance offered by the units constraints gives \PhySO\ an edge over other methods for finding the exact symbolic solutions, improving performance from a purely predictive standpoint. This makes \PhySO\ a potentially useful tool for opening up black-box physics models such as neural networks fitted on data of physical phenomena. In addition, we note that in the context of physics, components of our \PhySO\ framework can not only be used to improve the performance of algorithms built upon \cite{PetersenDSR}'s framework \citep{uDSR_DSRbased, SR_PG_improvements, SR_phy_demo_hamiltonian_DSRbased, Du_PDE_SR_DSRbased}, but can also be used in tandem with other approaches. For instance, our \insitu units prior can be used to reduce search space in the context of probabilistic or exhaustive searches \citep{Exhaustive_SR, Exhaustive_SR_wconstraints, Grammar_prior_wMC_SR, BSR_Bayesian_SR}, by severing physically impossible symbolic links in neuro-symbolic approaches \citep{Martius_SymbolsInNN, Sindy_SymbolsInNN, Zheng_SymbolsInNN_DSRbased, EQL_SymbolsInNN, Valle_SymbolsInNN, Kim_SymbolsInNN, Panju_SymbolsInNN}, during the seeding or mutation phases of genetic programming algorithms \citep{EureqaPaper2009, EureqaPaper2011_AFP, ITEA, EPLEX, FEAT, SBP_GP, PySR, Cranmer2020_GNN, GP_GOMEA, gplearn, OPERON, uDSR_DSRbased} or for making a physically motivated dataset of expressions, which in conjunction with enabling the RNN to be informed of local units constraints, could improve the performance of supervised approaches \citep{Kamienny_EndToEndSR, NeSymReS_EndToEndSR, Biggio_EndToEndSR, SymFormer_EndToEndSR, Becker_PDE_SR_EndToEnd, Kamienny_learning_mutations, uDSR_DSRbased}.\\


We recognize that in its current form, \PhySO\ needs to be provided the physical units of the free parameters it is allowed to use. Although this is typically not an issue for SR problems that tend to fall on the more theoretical side as constants that can appear in expressions if any are usually well known, in scenarios of novel empirical scientific exploration, the appropriate selection and units of free parameters may not be immediately evident. In such scenarios, we suggest the inclusion of one free parameter for each variable, matching their units. This approach grants \PhySO\ the flexibility to combine these parameters, or a subset thereof, to derive the most coherent combination that seamlessly integrates into the expression from a units perspective. As detailed in Appendix \ref{appendix:const_discovery}, utilizing this protocol enables \PhySO\ to accurately deduce formulae and the physical constants appearing in those. Examples include the recovery of expression for the terminal velocity during free fall, and its proportionality with the square root of an acceleration, by adeptly combining a velocity with an area to derive the acceleration parameter. In other examples, we show that \PhySO\ is able to effectively rediscover the universal gravitational constant or the ideal gas constant along with their units in addition to the expressions they intervene in.

Arguably, permitting a multitude of free parameters of various physical units, could inadvertently expand the search space. While this is a valid observation, it is worth noting that the algorithm remains significantly constrained, both by the limited assortment of these parameters and by the inherent units constraints between input variables, especially when considering dimensionless operations like $\cos, \exp$ and so forth. Moreover, given that the algorithm combines parameters based on the units of the variables and prioritizes solutions of lower complexity, the units of new physical constants typically align closely with the family of units of the problem, rather than assuming arbitrary values. Finally, it is worth noting that in addition to dimensional analysis constraints, another key finding of our study is that making the neural network able to observe units of symbols and currently required units in partially written expressions while they are being generated typically improves the recovery rate even without enforcing constraints directly.
However, resolving SR problems without knowing \emph{a priori} the units of the free parameters that can appear in the expressions is typically more difficult. We acknowledge this limitation and are actively considering future enhancements to \PhySO\ that would enable it to intelligently and autonomously ascertain the units of its free parameters.\\


Our approach is based on a deep reinforcement learning methodology, where the neural network is reinitialized at the start of each SR task. It is therefore trained independently for each specific problem, and so does not benefit from past experience nor is it pre-trained on a dataset of well known physical functional forms. One could argue that this makes our approach in principle ``unbiased'' akin to unsupervised learning setups and therefore well suited for discovering new physics \citep{ML_for_new_physics}. However, this also intrinsically limits SR capabilities as exploiting such prior knowledge is of great value for resolving the curse of accuracy guided SR described above. One can exploit such prior knowledge by formulating it as an \insitu prior \citep{DSR_wikipedia_prior, bayesian_prior} or by learning on it in a supervised manner using transformers learning techniques \citep{Kamienny_EndToEndSR, Kamienny_learning_mutations, Kamienny_learning_wassumptions, NeSymReS_EndToEndSR, SymFormer_EndToEndSR}. However, although state-of-the-art supervised SR methods, as of now, shine in providing accurate approximations, they show poorer exact symbolic expression recovery rates than other methods (see \eg the performances of \texttt{NeSymReS} in Figure \ref{fig:feynman_benchmark} or the ablation study conducted in \citealt{uDSR_DSRbased}).

While the combination of supervised and reinforcement learning may seem promising, \cite{uDSR_DSRbased} demonstrated that such a combination offers only marginal enhancements in exact symbolic recovery. Nonetheless, in the age of large language models, there is potential to harness vast internet-scale knowledge (see \eg \citealt{SymbolicGPT}). By learning the association between data points and mathematical expressions in realistic scenarios, and aligning with domain-specific assumptions using supervised learning techniques, it is conceivable to integrate this knowledge into a reinforcement learning framework, as exemplified by \cite{MineDojo_minecraft}. This approach might allow the recovery of expressions of substantially greater complexity than those we have explored in Section \ref{sec:case_studies}. In addition, while our approach generates a Pareto front that gives accuracy-complexity trade-offs, future enhancements that integrate both complexity and accuracy into a singular metric (as in \cite{Exhaustive_SR}) could potentially enhance symbolic regression performances and address model selection challenges.


As we have shown in Section~\ref{subsec:case_galactic}, it is straightforward to improve our method by combining it with the powerful problem simplification schemes devised in \citep{AIFeynman, AIFeynman2, Divide_and_conquer, GSR, Cranmer2020_GNN}. The results of the separability procedures implemented in the \citep{AIFeynman2} algorithm are conveniently recorded in separate datafiles, which makes it completely straightforward to use their approach as a pre-processing step for \PhySO. We anticipate that integrating their method within our algorithm, following the approach of \cite{uDSR_DSRbased}, should enhance the performance of \PhySO.

\section{Conclusions}
\label{sec:conclusions}

We have presented a new symbolic regression algorithm, built from the ground up to make use of the highly restrictive constraint that we have in the physical sciences that our equations must have balanced units. The heart of the algorithm is an embedding that generates a sequence of mathematical symbols while cumulatively keeping track of their physical units. We adopt the very successful deep reinforcement learning strategy of \citet{PetersenDSR}, which we use to train our RNN to not only produce accurate expressions but physically sound ones by making it learn local units constraints.

The algorithm was benchmarked and compared to 17 other baseline symbolic regression approaches on 120 cases from the Feynman Lectures on Physics and other textbooks. The results demonstrated the usefulness of constraints arising from dimensional analysis compared to \citet{PetersenDSR}, a purely reinforcement learning based baseline approach. In addition our approach achieved state-of-the-art leading performances in the presence of even minimal levels of noise (exceeding 0.1\%) and showing consistent performances up to $10 \%$ noise levels.

The algorithm was applied to several test cases from astrophysics. The first was a simple search for the energy of a particle in Special Relativity (Section~\ref{subsec:case_relativity}), which our algorithm was able to find, yet is a problem that the standard \cite{PetersenDSR} code fails on. The second test case applied the algorithm to the famous Hubble diagram of supernovae of type Ia. While the form of the Hubble parameter $H(z)$ in standard $\Lambda$CDM cosmology was indeed recovered, the algorithm finds that other simpler solutions fit the supernova data (in isolation) better. This result is consistent with the findings of \citet{Exhaustive_SR}. Another test examined a relatively complicated function in galactic dynamics, where we searched for the functional form of the radial action coordinate in an isochrone stellar potential model. This is an equation that neither the \citet{PetersenDSR} nor the \citet{AIFeynman2} methods are able to find. Although our algorithm initially fails in this test, we managed to recover the correct equation by first splitting the dataset using the additive separability criterion as implemented by \citet{AIFeynman}. 

These tests have demonstrated the applicability of the algorithm to model data of the real world as well as to derive non-obvious analytic expressions for properties of perfect mathematical models of physical systems. Although we realise that the physical laws potentially discovered by our method will depend on data range, choice of priors, etc, this is a step toward a full agnostic method for connecting observational data to theory. Future contributions in this research program will extend the algorithm to allow for differential and integral operators, potentially permitting the solution of ordinary and partial differential equations with physical units constraints. However, our primary goal will be to use the new machinery to discover as yet unknown physical relationships from the state-of-the-art large surveys that the astrophysical community has at its disposal.

\section*{Code availability}
\label{sec:availability}

The documented code for the \PhySO\ algorithm along with demonstration notebooks are available on GitHub \href{https://github.com/WassimTenachi/PhySO}{github.com/WassimTenachi/PhySO} \github{https://github.com/WassimTenachi/PhySO} with a frozen version related to this work deposited on zenodo: \href{https://doi.org/10.5281/zenodo.8415435}{10.5281/zenodo.8415435}.

\section*{Acknowledgments}
RI acknowledges funding from the European Research Council (ERC) under the European Unions Horizon 2020 research and innovation programme (grant agreement No. 834148).
The authors would like to acknowledge the High Performance Computing Center of the University of Strasbourg for supporting this work by providing scientific support and access to computing resources. Part of the computing resources were funded by the Equipex Equip@Meso project (Programme Investissements d'Avenir) and the CPER Alsacalcul/Big Data.

\bibliography{SymbolicRegression}
\bibliographystyle{aasjournal}

\appendix

\section{Datasets for the astrophysical examples}
\label{appendix:astro_data}

\def\mColored{\textcolor{myred}{m}}
\def\vColored{\textcolor{myred}{v}}
\def\cColored{\textcolor{myddblue}{c}}
\def\RelatEnergy{{$E = \frac{\mColored \cColored^2}{\sqrt{1 -\vColored^2/\cColored^2}}$\ }}
\def\EColored{\textcolor{myred}{E}}
\def\LColored{\textcolor{myred}{L}}
\def\GMColored{\textcolor{myddblue}{G M}}
\def\bColored{\textcolor{myddblue}{b}}
\def\IsochroneAction{{$J_r = \frac{\GMColored}{\sqrt{-2 \EColored}} - \frac{1}{2} \Big( \LColored + \frac{1}{2} \sqrt{\LColored^2 - 4 \GMColored \bColored} \, \Big)$\ }}
\def\rColored{\textcolor{myred}{r}}
\def\rhoColored{\textcolor{myddblue}{\rho_0}}
\def\RsColored{\textcolor{myddblue}{R_s}}
\def\NFWprofile{{$\rho = \rhoColored / \Big( \frac{\rColored}{\RsColored} (1 + \frac{\rColored}{\RsColored})^2 \Big)$\ }}
\def\tColored{\textcolor{myred}{t}}
\def\alphahoColored{\textcolor{myddblue}{\alpha}}
\def\PhiColored{\textcolor{myddblue}{\Phi}}
\def\fColored{\textcolor{myddblue}{\omega}}
\def\DHO{{$y = e^{-\alphahoColored \tColored} \cos (\fColored \tColored + \PhiColored) $\ }}
\def\moColored{\textcolor{myred}{m_1}}
\def\mtColored{\textcolor{myred}{m_2}}
\def\rColored{\textcolor{myred}{r}}
\def\GColored{\textcolor{myddblue}{G}}
\def\ClassicalGravity{{$F = \frac{\GColored \moColored \mtColored}{\rColored^2}$\ }}
\def\xColored{\textcolor{myred}{x}}
\def\OmegaColored{\textcolor{myddblue}{\Omega_m}}
\def\HColored{\textcolor{myddblue}{H_0}}
\def\ExpansionLaw{{$H^2(x\equiv1+z) = \HColored^2 (\OmegaColored \xColored^3 + (1 - \OmegaColored))$\ }}

This appendix gives details regarding the synthetic datasets for the astrophysical examples. For each case, we generate 1000 noiseless data points following a random uniform law using arbitrary scales for the mock data. Table \ref{table:astro_expressions} gives the target expressions and Table \ref{table:astro_input_vars} and \ref{table:astro_constants} give details regarding the variables and constants appearing in those expressions.

\begin{table*}[h]
\begin{center}
\begin{tabular}{|l@{\hskip 0.5in}l|}
Case                       & Expression          \\ \hline
Relativistic Energy        & {\RelatEnergy}      \\
Isochrone Action           & {\IsochroneAction}  \\
NFW Profile                & {\NFWprofile}       \\
Damped Harmonic Oscillator & {\DHO}              \\
Classical Gravity          & {\ClassicalGravity} \\
Expansion Law              & {\ExpansionLaw}    
\end{tabular}
\end{center}
\caption{Astrophysical examples target expressions. Input variables and free parameters choosable by \PhySO\ as symbols are colored in red and blue respectively, with fixed constants left in black.}
\label{table:astro_expressions}
\end{table*}

\begin{table*}[h]
\begin{center}
\begin{tabular}{|lc|ccc|ccc|ccc|}
\multicolumn{2}{|c|}{Output}               & \multicolumn{3}{c|}{Variable 1}           & \multicolumn{3}{c|}{Variable 2}   & \multicolumn{3}{c|}{Variable 3} \\ \hline
\multicolumn{1}{|c}{Name} & Units          & Name  & Range              & Units        & Name  & Range    & Units          & Name     & Range     & Units    \\ \hline
$E$                       & $M.L^2.T^{-2}$ & m     & [-10,10]           & $M$          & v     & [-9,9]   & $L.T^{-1}$     &          &           &          \\
$J_r$                     & $L^2.T^{-1}$   & L     & [2.3, 3]           & $L^2.T^{-1}$ & E     & [-4, -6] & $M.L^2.T^{-2}$ &          &           &          \\
$\rho$                    & $M.L^{-3}$     & r     & [0.2, 3]           & $L$          &       &          &                &          &           &          \\
$y$                       & 1              & t     & [$1.5\pi$, $7\pi$] & $T$          &       &          &                &          &           &          \\
$F$                       & $M.L.T^{-2}$   & $m_1$ & [0,1]              & $M$          & $m_2$ & [0,1]    & $M$            & r        & [1,4]     & $L$      \\
$H^2$                     & $T^{-2}$       & z     & [0.01, 2.5]        & 1            &       &          &                &          &           &         
\end{tabular}
\caption{Data range and units of the output and input variables appearing in the astrophysical examples.}
\label{table:astro_input_vars}
\end{center}
\end{table*}

\begin{table*}[h]
\begin{tabular}{|ccc|ccc|ccc|}
\multicolumn{3}{|c|}{Constant 1}    & \multicolumn{3}{c|}{Constant 2} & \multicolumn{3}{c|}{Constant 3} \\ \hline
Name  & Value & Units               & Name      & Value  & Units      & Name      & Value    & Units    \\ \hline
c     & 10    & $L.T^{-1}$          &           &        &            &           &          &          \\
GM    & 0.467 & $L^3.T^{-2}$        & b         & 1.234  & $L$        &           &          &          \\
$r_s$ & 1.391 & $L$                 & $\rho_0$  & 0.984  & $M.L^{-3}$ &           &          &          \\
$\omega$     & 0.784 & $T^{-1}$            & $\alpha$  & 0.101  & 1          & $\phi$    & 0.997    & 1        \\
G     & 1.184 & $L^3.M^{-1}.T^{-2}$ &           &        &            &           &          &          \\
$H_0$ & 1.072 & $T^{-1}$            & $\Omega$  & 1.315  & 1          &           &          &         
\end{tabular}
\caption{Target value and units of constants appearing in the astrophysical examples.}
\label{table:astro_constants}
\end{table*}

\pagebreak

\section{Discovering both analytical laws \& constants of Nature}
\label{appendix:const_discovery}

\def\nColo{\textcolor{myred}{n}}
\def\TColo{\textcolor{myred}{T}}
\def\VColo{\textcolor{myred}{V}}
\def\GasLaw{{$P = \frac{\nColo R \TColo}{\VColo}$\ }}

\def\mColo{\textcolor{myred}{m}}
\def\rhoColo{\textcolor{myred}{\rho}}
\def\AColo{\textcolor{myred}{A}}
\def\TerminalVelocity{{$v_t = \sqrt{\frac{2 \mColo g}{\rhoColo \AColo C_d}}$\ }}

\def\moColored{\textcolor{myred}{m_1}}
\def\mtColored{\textcolor{myred}{m_2}}
\def\rColored{\textcolor{myred}{r}}
\def\GColored{\textcolor{myddblue}{G}}
\def\ClassicalGravityFree{{$F = \frac{G \moColored \mtColored}{\rColored^2}$\ }}

\def\nuColo{\textcolor{myred}{\nu}}
\def\TColo{\textcolor{myred}{T}}
\def\PlanckLawPhotons{{$n = {1}/({e^{\frac{h \nuColo}{k_b \TColo}}-1})$\ }}

\def\EaColo{\textcolor{myred}{E_1}}
\def\EbColo{\textcolor{myred}{E_2}}
\def\DPhi{\textcolor{myred}{\Delta\Phi}}
\def\WaveInterference{{$E = \EaColo + \EbColo + 2\sqrt{\EaColo \EbColo} + \cos{\DPhi}$\ }}

We note that for new scientific discovery, there are instances where the appropriate free parameters and their corresponding units are not immediately evident. In such situations, we propose a protocol wherein \PhySO\ is allowed one free parameter for each input variable, sharing the same units, and another free parameter reflecting the units of the output variable. Specifically, for an SR problem consisting in the deduction of $y$ from $\{x_1, ... x_n \}$, we would permit the inclusion of $\{\theta_{y}, \theta_{x_1}, ... \theta_{x_n} \}$ as free constants. This grants \PhySO\ the flexibility to selectively combine or omit these free parameters to construct new parameters that align with dimensional analysis constraints. In light of these combinations, we adjust the center of the soft length prior to a length of 12, facilitating longer expressions.\\

In this more demanding setup, we demonstrate that \PhySO\ can adeptly resolve the SR challenges outlined in Table \ref{table:free_expressions} (with dataset details given in Table \ref{table:free_input_vars} and \ref{table:free_constants}), yielding both the precise symbolic expressions and their corresponding physical constants with accurate units. The scripts employed for these experiments are accessible in our repository.\\

\begin{table*}[h]
\begin{center}
\begin{tabular}{|l@{\hskip 0.5in}l|}
Case                        & Expression              \\ \hline
Ideal Gas Law               & {\GasLaw}               \\
Free Fall Terminal Velocity & {\TerminalVelocity}     \\
Classical Gravity           & {\ClassicalGravityFree} \\
Black Body Photon Count     & {\PlanckLawPhotons}     \\
Wave Interference           & {\WaveInterference}    
\end{tabular}
\end{center}
\caption{Target expressions. Input variables are colored in red.}
\label{table:free_expressions}
\end{table*}

\begin{table*}[h]
\begin{center}
\begin{tabular}{|cc|ccc|ccc|ccc|}
\multicolumn{2}{|c|}{Output} & \multicolumn{3}{c|}{Variable 1}  & \multicolumn{3}{c|}{Variable 2}  & \multicolumn{3}{c|}{Variable 3}   \\ \hline
Name    & Units              & Name  & Range   & Units          & Name   & Range  & Units          & Name         & Range    & Units   \\ \hline
P       & $L^{-1}.T^{-2}.M$  & n     & [1,5]   & $N$            & T      & [1,5]  & $\Theta$       & V            & [1,5]    & $L^3$   \\
$v_t$   & $L.T^{-1}$         & m     & [1, 10] & $M$            & $\rho$ & [1, 6] & $M.L^{-3}$     & A            & [1, 5]   & $L^{2}$ \\
F       & $L.T^{-2}.M$       & $m_1$ & [1, 5]  & $M$            & $m_2$  & [1, 5] & $M$            & r            & $[1, 5]$ & $L$     \\
n       & 1                  & $\nu$  & [1, 5]  & $T^{-1}$       & T      & [1, 5] & $\Theta$       &              &          &         \\
E       & $L$                & $E_1$ & [1, 5]  & $L^2.T^{-2}.M$ & $E_2$  & [1, 5] & $L^2.T^{-2}.M$ & $\Delta\Phi$ & [-5, 5]  & 1  
\end{tabular}
\caption{Data range and units of the output and input variables appearing in the examples.}
\label{table:free_input_vars}
\end{center}
\end{table*}

\begin{table*}[h]
\begin{center}
\begin{tabular}{|ccc|ccc|}
\multicolumn{3}{|c|}{Constant 1}                    & \multicolumn{3}{c|}{Constant 2}              \\ \hline
Name & Value & Units                                & Name  & Value & Units                        \\ \hline
R    & 8.314 & $L^{-2}.T^{-2}.M.N^{-1}.\Theta^{-1}$ &       &       &                              \\
g    & 9.807 & $L.T^{-2}$                           & $C_d$ & 0.470 & 1                            \\
G    & 6.674 & $L.T^{-2}.M$                         &       &       &                              \\
h    & 6.626 & $L^{2}.T^{-1}.M$                     & $k_b$ & 1.123 & $L^{2}.T^{-2}.M.\Theta^{-1}$ \\
-    & -     & -                                    &       &       &                            
\end{tabular}
\caption{Target value and target units of constants appearing in the examples.}
\label{table:free_constants}
\end{center}
\end{table*}

For illustration, \PhySO\ successfully derives the equation describing the equation of state of an ideal gas $P=C\frac{n T}{V}$ with $C=\frac{\theta_{P} \theta_{V}}{\theta_{n} \theta_{T}}$ having units $M.L^2.T^{-2}.K^{-1}.N^{-1}$ effectively rediscovering the ideal gas constant usually denoted by $R$.
Similarly, \PhySO\ is able to recover the expression for the terminal velocity of a free falling object as a function of its mass $m$, its surface area $A$ and the density of the medium it traverses $\rho$ as $v_t = \sqrt{C \frac{m}{\rho A}}$ by unveiling its proportionality to the square root of an acceleration $\sqrt{C}$, formulated by \PhySO\ as $\sqrt{{\theta_{v_t}}/{\sqrt{\theta_{A}}}}$, corresponding to the Earth surface gravity $\sqrt{g}$ and other scale factors.
Furthermore, \PhySO\ identifies the gravitational force in relation to the involved masses $m_1$, $m_2$ and distance $r$ as $F = C {m_1 m_2}/{r^2}$ discovering the need for a constant $C$ having units $L^3.T^{-2}.M$ formulated by \PhySO\ as $C = {\theta_F {\theta_r}^2}/{{\theta_{m_1}}^2}$, effectively rediscovering the gravitational constant $G$ in the process.
In an other scenario, deriving the number density of photons recovered from a black body at any given temperature $T$ and frequency $\nu$, \PhySO\ is able to recover $n = 1/(e^{\nu C/ T}-1)$ where $C$ represents the quotient $C=h/k_b$, $h$ and $k_b$ denoting the Planck and Boltzmann constants, respectively.
In most aforementioned cases, \PhySO\ judiciously combined a subset of the available free parameters to pinpoint the precise constants needed to resolve the SR problems through a physically consistent physical law. In this last example, we show that \PhySO\  recognizes scenarios where free parameters are largely redundant as it is able to derive the energy $E$ resultant from the interference of two waves, given their energies $E_1$, $E_2$ and their phase shift $\Delta\Phi$ without the need for any of $\{\theta_{E}, \theta_{E_1}, \theta_{E_2}\}$.\\

\end{document}